\documentclass[manuscript]{ar2e1}
\usepackage[margin=1in]{geometry}

\def\msun{\hbox{M$_\odot$}}
\def\zsun{\hbox{Z$_\odot$}}

\def\hauv{\hbox{H$\alpha$-to-UV}}
\def\mhat{\hbox{M$_{\rm c}$}}
\def\lm{\hbox{L$_{\rm V}$/M$_{\rm dyn}$}}
\def\lmpop{\hbox{L$_{\rm V}$/M$_{\rm pop}$}} 
\def\mdyn{\hbox{M$_{\rm dyn}$}}
\def\mpop{\hbox{M$_{\rm pop}$}}

\def\mc{\hbox{M$_{\rm c}$}}
\def\mstar{\hbox{M$_{\star}$}}
\def\dave{\hbox{M$_{\star}$-SFR}}

\begin{document}

\input epsf.tex    
\input epsf.def   

\input psfig.sty

\jname{..}
\jyear{2000}
\jvol{}
\ARinfo{1056-8700/97/0610-00}

\title{A Universal Stellar Initial Mass Function?  A Critical Look at Variations }

\markboth{IMF variations?}{IMF Universality?}

\author{Nate Bastian$^1$,  Kevin R. Covey$^{2,3}$, Michael R. Meyer$^{4,5}$
\affiliation{$^1$ Institute of Astronomy, University of Cambridge, Madingley Road, Cambridge CB3 0HA, UK\\
$^2$ Department of Astronomy, Cornell University, Ithaca, NY 14853, USA\\
$^3$ Harvard Smithsonian Center for Astrophysics, 60 Garden St., Cambridge, MA 02138, USA\\
$^4$ Institute of Astronomy, ETH Z\"{u}rich, Wolfgang-Pauli-Str. 27, 8093 Z\"{u}rich, Switzerland\\
$^5$ Steward Observatory, The University of Arizona, Tucson, AZ 85721, USA\\
}}


\begin{abstract}

Few topics in astronomy initiate such vigorous discussion as whether or not the initial mass function (IMF) of stars is universal, or instead sensitive to the initial conditions of star formation.  The distinction is of critical importance: the IMF influences most of the observable properties of stellar populations and galaxies, and detecting variations in the IMF could provide deep insights into the process by which stars form.  In this review, we take a critical look at the case for IMF variations, with a view towards whether other explanations are sufficient given the evidence.  Studies of the field, local young clusters and associations, and old globular clusters suggest that the vast majority were drawn from a ``universal" IMF: a power-law of Salpeter index ($\Gamma=1.35$) above a few solar masses, and a log normal or shallower power-law ($\Gamma \sim 0-0.25$) between a few tenths and a few solar masses (ignoring the effects of unresolved binaries).  The shape and universality of the IMF at the stellar-substellar boundary is still under investigation and uncertainties remain large, but most observations are consistent with a IMF that declines ($\Gamma < -0.5$) well below the hydrogen burning limit.  Observations of resolved stellar populations and the integrated properties of most galaxies are also consistent with a ``universal IMF", suggesting no gross variations in the IMF over much of cosmic time.  
There are indications of ``non-standard" IMFs in specific local and extragalactic environments, which clearly warrant further study.  
Nonetheless, there is no clear evidence that the IMF varies strongly and systematically as a function of initial conditions after the first few generations of stars.   We close with suggestions for future work that might uncover more subtle IMF variations than those that could be discerned to date. 

\end{abstract}

\maketitle

\section{Introduction}

As every undergraduate astronomy student knows,
a star's mass (and, to a lesser degree, its chemical composition),
determines its subsequent evolutionary path. 
In some sense, the important question for stellar evolution becomes:
how are stellar
masses determined?  This is also one of the central questions in developing
a theory of star formation.

Foremost among the tools we have to investigate the origin of stellar masses is
the initial mass function (IMF).  The IMF was first introduced by Salpeter (1955) 
and provides a convenient way of parameterizing the relative numbers of 
stars as a function of their mass.  In addition to informing our understanding of 
stellar origins and evolution, the IMF is an important input to many
astrophysical studies. Indeed, most problems in modern
astrophysics can be ``solved" by invoking an IMF that varies in some specific
way. By changing the IMF as a function of cosmic time, for example,
one can derive star formation histories of the Universe that turnover,
remain flat, or even continue to rise to very high redshift.  The explanatory power of these ``solutions", however, is rather limited without independent observational evidence of IMF variations.  

Where we do have observations of the IMF, however, any predictive theory
of star formation must explain its shape, as well as how it might vary with
initial conditions.  Turned the other way round, we can ask:  is the
IMF universal
or does its shape vary?  If we can confidently observe variations in
the IMF, we can hope to study those variations to understand the critical scales
or conditions under which stars of a certain mass form, informing a
quantitative
prescription for star formation (e.g. McKee \& Ostriker~2007).


\subsection{Functional Forms}

In his famous 1955 paper, E. Salpeter introduced a power law IMF of the form
\begin{equation}
\Phi(logm) = dN/d\log m \propto m^{-\Gamma},
\label{eqn:salpeter}
\end{equation}

\noindent where $m$ is the mass of a star and N is the number of stars in some logarithmic mass range $log m+d log m$.
Integrating this function and deriving a proper normalization, we can
calculate the number of stars within a logarithmic mass interval.
Hereafter we will refer to single--power law IMFs as ``Salpeter-like
IMFs''; those IMFs with $\Gamma \sim 1.35$ we will refer to as having
a ``Salpeter slope'' (note that by ``slope" we are referring to the ``logarithmic slope" throughout this review).

It was recognized in the late 1970s that the IMF was probably not a
single power law over all stellar masses.  Through the 1980s  and
early 1990s various alternatives were explored, including a
multi-segment power law (Kroupa, Tout \& Gilmore 1993) where the slope
of the IMF at lower masses was found to be shallower than the Salpeter
value obtained at higher masses.  Hereafter we will refer to these
segmented power--law IMFs as ``Kroupa IMFs''.

One can also discuss the IMF in the context of a mass spectrum,
formulating a distribution function in linear mass units
\begin{equation}
\chi(m) = dN/dm \propto m^{-\alpha},
\label{eqn:powerlaw}
\end{equation} enabling an estimation of the number of  stars within
some mass range.  Notice that
\begin{equation}
\chi (m) = 1/(m \ln10) \Phi(\log~m)
\end{equation}
yielding the relation
\begin{equation}
\alpha = \Gamma + 1.
\end{equation}
Quoting (and misquoting) results for the ``slope"  of the IMF in linear
rather than logarithmic units (or vice versa) has led to enormous
confusion in the literature.  The numerical values of $\Gamma$ and
$\alpha$ are often misinterpreted, but perhaps more pernicious are the
subtle implications of confusing less precise descriptions of the mass
function slope: an IMF that is ``flat'' or ``falling''  in logarithmic
units can still be rising in linear units!   A key reference point in
this discussion is the relative amount of mass contained within
equally sized logarithmic mass bins.  Taking the first moment of the
distribution described by Equation \ref{eqn:powerlaw} with respect to
mass, we can calculate the total mass within a logarithmic interval.  The
critical slope is  $\alpha > 2$ ($\Gamma >1$) where the amount of
mass per equal sized logarithmic bin is larger for the lower mass bins.

The logarithmic formalism of the IMF is preferred by some researchers for
two reasons:  1) it provides a quick estimate of the relative stellar
mass in decadal bins and 2) it permits description of the IMF as a
log-normal function.  The log-normal form of the IMF was first
introduced by Miller \& Scalo (1979) and a theoretical explanation was
offered by Zinnecker~(1984; see also Larson~1973) who invoked the central limit theorem.
The central limit theorem states that any function resulting from the
sum of an infinite number of independent variables can be described by
a normal or gaussian distribution function. If we imagine star
formation as a complex transformation where stellar mass is determined
by the product of several variables and then take the log, we have
that log-m is the sum of a series of possibly independent distribution
functions.  If the series is infinite and the variables are truly
independent, we can then expect the distribution of log-m to follow a
gaussian form.  In short, if star formation is very complex, it would
not be a surprise if the distribution of log-m were gaussian, i.e.
log-normal  (see Adams \& Fatuzzo, 1996) and have the form
\begin{equation}
\phi(m) \sim e^{-\frac{({\log}~m-{{\log}~m_c})^2}{2\sigma^2}}.
\label{eqn:lognormal}
\end{equation}

\begin{figure}[!h]	
\centerline{\psfig{figure=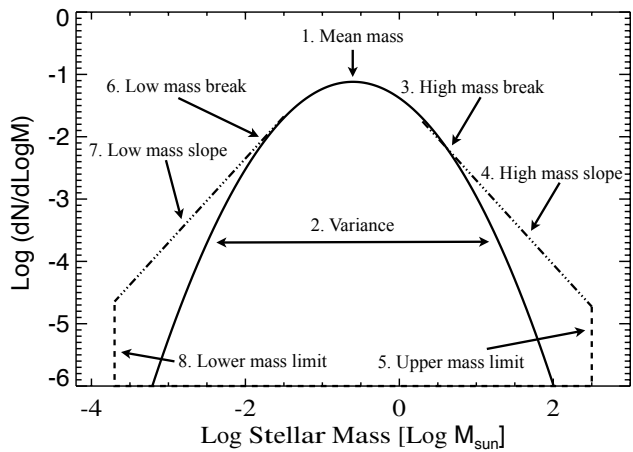,height=20pc,angle=270}}
\caption{Schematic of an eight parameter IMF.  The ``base" of the IMF
is approximated as a log-normal distribution (shown as a solid line) with a 1) characteristic mass and
2) variance.  On the high mass side one has the additional parameters
of the 3) high mass break, 4) high mass slope (shown as a dashed-dotted line) and 5) upper mass limit (represented by a dashed line).
 The 6th, 7th and 8th parameters are the equivalent on the low mass
end.}
\label{fig:imf}
\end{figure}

Hereafter we refer to the variable $m_c$ as the characteristic mass in a log--normal IMF.  A log--normal function is shown as a solid line in
Fig.~\ref{fig:imf} as the base of the IMF.  It is useful to note that
the power--law form of the IMF ($dN/dlogm$) is a straight line in this
log--log plot.  If we imagine that the Salpeter law holds above some
limit (high mass break), and that below this limit the IMF is log-normal, we can
describe the IMF with four parameters:  characteristic mass, dispersion,
power-law slope, and break point between the two distributions.  In
fact, this description appears to be a useful one for many
applications:  Chabrier (2003) reviewed numerous observational constraints and presented a new formulation of the log-normal IMF\footnote{Dabringhausen, Hilker \& Kroupa~(2008) have shown that the Chabrier IMF is extremely similar to a two-part power-law, hence distinguishing between a Kroupa-type or Chabrier-type IMF is very difficult.}.  

Detailed studies of the IMF over the full stellar/sub-stellar mass range, however, may require an eight parameter description of the IMF.  In addition
to the four parameters described above, we have the reflection of the
high mass parameters for the low mass end, namely the break point
between the log-normal and  a low-mass power-law slope and the value
for this slope.  Finally, to ``cap" the ends of the distribution, we
need to introduce the lower and upper limits, both being subjects of
great debate.  

An additional parameterization of the IMF that is quite practical is
that of a ``tapered power-law" (de Marchi, Paresce \& Portegies
Zwart~2005).  This has the form
\begin{equation}
\chi(m) = \frac{dN}{dm} \propto m^{-\alpha} [1 - e^{(-m/m_{p})^{\beta}}]
\label{eqn:tapered}
\end{equation}
where $m_p$ is the peak mass (similar to m$_c$ in a Equation~\ref{eqn:lognormal}), $\alpha$ is the index of the power-law in the upper end
of the mass function (similar to $\alpha$ in Eqn.~\ref{eqn:powerlaw}) and
$\beta$ is the ``tapering" exponent to describe the lower end of the
IMF.  

\subsection{Observational Challenges}

Aside from the conceptual difficulty of adopting a succinct
mathematical description of the IMF, there are significant challenges
that must be overcome to provide a robust empirical measurement of the
IMF in various environments.  Scalo (1986) provides an excellent
introduction to the methodology used in constructing the IMF and we will
simply highlight the issues most relevant to our
discussion of IMF measurements in various environments.

\subsubsection{Star Formation History and Multiplicity}
\label{sec:sfh_binaries}

A primary motivation for many IMF studies is to search for IMF variations. Ironically, many of these studies require the implicit assumption that the IMF is constant over some spatial or temporal scale.  
Measuring the IMF of a complex, multi-component stellar population like the 
Milky Way disk  provides a useful case study.  The first step appears straight--forward: count stars within a limited volume as a function of mass.  Low mass stars are more numerous in the Galactic field than high
mass stars, however, so one must use large volumes to construct samples with significant numbers of high mass stars.  The
volume density of stars in each sample can be appropriately normalized, but
constructing one IMF from these data requires the assumption that
the IMF does not vary between those two volumes.

The second implicit assumption is required due to
finite stellar lifetimes.  If one counts stars in a population
spanning a large range of ages (as in the disk of the Milky Way) one
can only construct the  present day mass function (PDMF).  In
this example, some fraction of the more massive stars with lifetimes
shorter than the age of the Galaxy ($<$ 10 Gyr) will have evolved off
the main sequence and are no longer present in the Galactic disk, and
thus will not be represented in the PDMF.  Every star with M$\le$0.8
M$_{\odot}$ (i.e. spectral type mid-K or later) that was ever formed is still present
in the Galactic disk, however, and thus will be represented in the
PDMF.  Correcting the PDMF for the loss of previous generations of
high mass stars therefore requires the assumption that the IMF does not
vary in time, as well as knowledge of the star formation history (SFH) of the
Milky Way!  Elmegreen \& Scalo~(2006) provide a detailed exploration of
the degeneracies involved in inferring a stellar population's IMF from its measured PDMF and adopted SFH.  It is somewhat disconcerting that the ``break" from the
Salpeter law identified in many IMF studies needs to be invoked near
the place where the correction becomes important (around one solar
mass).  

Knowledge of the star formation history of the Galaxy is also required to understand the opposite end of
the mass spectrum.  Sub--stellar objects never stabilize through
hydrogen burning, so their mass--luminosity relation
evolves strongly with time.  A brown dwarf's luminosity can therefore be 
reconciled with numerous degenerate combinations of mass and age, such that
measuring the sub--stellar IMF in the Galactic Field requires good source 
counts down to very faint magnitudes and a very detailed knowledge of the star formation history of the Galaxy (Reid et al. 1999).   

The last observational problem for determining the IMF that we will address is 
stellar (and sub--stellar) multiplicity.
Many stellar objects are found to have gravitationally bound
companions of lower mass (by conventional definition) and at a variety
of orbital separations.  If we consider that many determinations of
the IMF start with monochromatic luminosity functions, we can guess
that unresolved companions of a range of mass (and therefore added
luminosity)  could wreak havoc on the IMF.  Indeed they do
(Kroupa~2001; Chabrier~2003; Ma\'{i}z Apell\'{a}niz~2008), but correcting for this effect
is difficult, because the impact of multiplicity on IMF studies is sensitive
to both the binary fraction and the mass ratio of the systems involved.  It
is straightforward that a higher multiplicity fraction will produce a larger
multiplicity correction, but the mass ratio dependence is somewhat more subtle:
systems with high mass ratios are effective at ``hiding" stars, and are
also quite difficult to detect as a multiple system. By contrast, equal mass systems will ``hide" a
star, but the system's increased luminosity allows it to be detected over a larger volume,
which compensates to some extent for the missing companions in luminosity
functions constructed from the field population (Reid \& Hawley 2005).

Correcting for this effect requires knowledge of the mass dependence of the multiplicity fraction and mass ratio distribution.  Observations indicate that multiplicity declines with primary mass (e.g. Lada 2006): the initial binary
frequency of O stars may be as high as 100\% (Mason et al. 2009)
dropping to $\sim$60\% for solar-type stars (Duquennoy
\& Mayor 1991a,b),  $\sim$30\% for early M stars (Fischer \& Marcy
1992, Reid \& Gizis 1997, Leinert et al. 1997, Delfosse et al. 2004),
and to 20\% or less for very-low mass objects (M$<$0.1 M$_{\odot}$;
Bouy et al. 2003, Burgasser et al. 2007).  The mass ratio distribution appears similarly
mass dependent, with typical O star mass ratios near unity
(Zinnecker \& Yorke 2007), a flat distribution of (detectable) mass ratios between 0.1 and 1.0 for solar type stars, and a preference towards equal mass systems for the lowest mass binaries.  These multiplicity rates and mass ratio distributions can be used to crudely correct observed mass functions for missing detections of unresolved secondaries.  In the 0.1--1.0 \msun\ regime,
converting from a system IMF to a single star IMF typically corresponds
to a $\sim$0.2 steepening of a rising power-law slope (i.e. the
Salpeter distribution would change from $\Gamma=1.35$ to
$\Gamma=1.55$).  Corrections for stellar multiplicity require yet another 
implicit assumption, in this case that multiplicity properties are spatially and temporally constant.  Investigations of the mass function of companions is an interesting area of ongoing research (e.g., Metchev \& Hillenbrand~2009, Goodwin \& Kouwenhoven 2009).  For the purposes of this review, we are only concerned with differences between two IMF determinations: we do not attempt to determine whether those differences are due to changes in the system IMFs, the companion mass function, or both.

\subsubsection{Statistical Considerations for IMF Measurements of
Resolved Stellar Populations}
\label{sec:statistics}

Aside from the observational difficulties described above,
substantive claims about IMF variations require a statistical
assessment of the agreement (or lack thereof) between the IMF measured
in different environments.  Most authors conduct this analysis by fitting 
a power-law or log-normal IMF to their measurements, and then compare the parameters of their fit
to those derived in other IMF studies.  A disagreement between the
derived parameters is typically interpreted as indicative of
astrophysically meaningful IMF variations.

The statistical significance of disagreements between IMF parameters
derived in different studies can be difficult to infer, however.  Some
studies do not provide uncertainties associated with their analytic
fits, and fewer provide a quantitative justification of the functional
form they adopt to describe the IMF.  Functional forms with more free
parameters can better capture the structure in a given IMF
measurement, but that structure may not be statistically significant.
Moreover, applying different functional forms to a single 
derived IMF reveals that the functional form adopted
can influence the perceived character of the resultant 
``measurement" (e.g. Jeffries et al. 2004).  Statistical tests such as the F-test
can inform the selection of an appropriate analytic form for a given IMF measurement, indicating if a fit with more degrees of freedom produces a large enough improvement in the $\chi^2$ value to justify their inclusion (e.g. Covey et al. 2008).  Similarly, robust statistical techniques can be used to minimize biases in the output parameters (e.g. Maschberger \& Kroupa 2009). 

Even when using well motivated functional forms to describe an
empirical IMF, care must be taken in interpreting the derived 
parameters.  If the IMF varies smoothly
with mass, as with the log-normal characterization, the shape one
expects to measure will depend on the mass range over which the
measurement is made.  
This is demonstrated in Figure 2, where the
value of $\Gamma$ derived from a power-law fit to a log-normal IMF
varies linearly with log mass, from 1.0 at 1 \msun\ to $\sim -0.6$ at
0.1 \msun\ and $\sim -2$ at 0.01 \msun.  The parameters one derives
from an analytic IMF fit may also be sensitive to the manner in which the
data were binned (Ma\'{i}z Apell\'{a}niz \& Ubeda 2005; Cara \& Lister 2008).
Two IMF measurements could therefore be consistent with being drawn
from the same parent population, but produce very different analytic
fits depending on what range of masses they sample, and how the data
were binned.  Statistical tests such as the Kolmogorov-Smirnov test can 
offer additional insight by addressing whether two distributions
are consistent with having been drawn from the same underlying
parent population directly without the artifice of binning the data.
Of course the results of such tests can be sensitive to systematic
differences in how two datasets (in this case stellar masses)
were constructed.

\subsection{The Utility of the IMF}
\label{sec:intro_applications}

The IMF is used in countless ways to describe stellar populations from star clusters to galaxies.  In some of our nearest galactic companions (e.g.~those in the local group) the IMF is probed in much the same way as in the Milky Way.  However, crowding and sensitivity limit studies of the IMF to $>0.4$~\msun\ in the LMC/SMC with HST imaging and above one solar mass beyond.  Yet in the search for variations in the IMF, it is vital to sample the most extreme conditions where star formation could be very different. As such conditions are rare, we need to probe larger volumes to encounter these extremes in 
density, metallicity, star formation rate, radiation field, and other properties. 
Furthermore, the IMF is a very useful tool for studying the integrated properties 
of distant galaxies where we cannot resolve individual stars.

The observable properties of a galaxy (i.e., color and magnitude), or any part thereof, is determined in large part by its IMF and star-formation history.  To aid in the understanding of a myriad of extragalactic observations, stellar population synthesis models (e.g. Leitherer et al.~1999; Bruzual \& Charlot~2003; Anders \& Fritze-v. Alvensleben~2003; Maraston~2005) have been developed, which, when given an input IMF, can integrate evolutionary models as a function of star formation history to predict observed colors and luminosities.  Observations can then be used to constrain the inferred star formation histories as well as the mass of a given stellar population.  Specific observations (e.g. B-band luminosity, near-IR color, etc) often preferentially sample a given stellar mass range.  For all but the most massive stars, multiple epochs of star formation contribute to each mass range. Hence a detailed knowledge of the star formation history (and metallicity) of the population is needed if one wants to constrain the underlying IMF.  These degeneracies between star formation history (SFH), metallicity, and the IMF, will be even more important for unresolved systems, and will be highlighted throughout this review.

If the IMF varies systematically with environment or metallicity (both of which could depend on cosmic look--back time), then it is possible, even likely, that the inferred SFHs, stellar masses and hence stellar mass density estimates would be systematically in error.  This could strongly bias our understanding of many important topics, from chemical evolution to how galaxies are formed.  One important quantity often measured for local and high-redshift galaxies is their present star formation rates.  By constraining the amount of ionizing radiation emitted by a galaxy (traced by UV, $H\alpha$, or infrared luminosity and correcting for extinction as needed), one can estimate the number of high mass stars present.  Combined with estimates of their lifetimes, these data can then be used to estimate the rate at which massive stars are forming.  However, massive stars make up only a very small fraction (in number and mass) of all the stars formed, so one needs to understand the relative number of high to low mass stars (i.e. the IMF) to determine the total mass that is being formed in stars (cf. Kennicutt~1998).   Hence the assumed IMF has a large impact in converting observed properties (e.g. H$\alpha$ luminosities) to star formation rates.

\subsection{The Scope of This Review} 

In this paper, we do not attempt to review all work on the IMF.   The 
reader is referred to the classic works of Salpeter (1955), 
Miller \& Scalo (1979), and in particular Scalo (1986) for
a discussion of previous IMF studies. 
More recently, the excellent review by Chabrier (2003) synthesizes important
work in many areas and adopts for the log--normal form of the IMF.  A comprehensive overview of the methodology and results of IMF studies at the low mass end can also be found in "New Light on Dark Stars" (Reid \& Hawley~2005).  Some important updates can be found in "The IMF at 50" conference proceedings
celebrating 50 years since the publication of Salpeter's landmark paper. 
Elmegreen~(2009) also discusses many topics related to the IMF from an 
extragalactic point of view. 

Two important aspects of IMF studies that we will not cover in this review are: 1) the first generation of stars (i.e. Population~III stars) and 2) the implications of a universal or variable IMF on star formation theories.  For the former topic we refer the interested reader to the recent reviews of Bromm \& Larson~(2004), Glover~(2005) and Norman~(2008).  For the latter topic, we refer the reader to McKee \& Ostriker~(2007) and Zinnecker \& Yorke~(2007).

We will focus in this review on more recent determinations of the IMF, with 
particular attention to claims of non-standard/varying IMFs.  
Of course any two finite samples drawn from a distribution function will have
small differences due to Poisson statistics.  We consider these ``variations''
trivial and will concern ourselves mainly with observations where the inferred
IMF is reported to be inconsistent with having been drawn from the same
distribution function that is thought to characterize the IMF elsewhere
(i.e. with a low probability that two IMFs were drawn from the same
parent population - see Elmegreen~(1999) and Gilmore~(2001) for discussions on stochasticity in the IMF).  
While investigating claims of non-universal IMFs, we ask 
the question:  is an unusual IMF required by the observations or 
is another more pedestrian explanation suitable?

\section{Stellar Populations in the Galactic Disk}
\label{sec:pop1}

We begin by considering IMFs measured for various stellar populations within the disk of the Milky Way: volume-limited samples of stars at the solar circle in the Milky Way, otherwise known as ``the field" (\S \ref{sec:field});  open clusters (simple stellar populations of fixed age and metallicity, \S \ref{sec:open_clusters}); nearby regions of active star formation (\S \ref{sec:nearby_young_clusters}); and young star clusters representing the extremes of star formation in parameters such as cluster mass, star formation rate, and galactic potential (\S 2.4).  The bulge of the Galaxy and globular clusters will be discussed in \S~\ref{sec:near-field}.

\subsection{The Field Star IMF}
\label{sec:field}

Nearby stars with high quality trigonometric parallax measurements are a 
valuable sample from which to measure a stellar IMF.  Imaging surveys
of these stars provide the most detailed census of stellar
multiplicity, and thus the purest measurement of the single star IMF
and the multiplicity correction to be applied to system IMFs measured
in more distant environments.  As the mass of an isolated star can
only be inferred from its luminosity (adopting a metallicity dependent
mass-luminosity relation), distance determinations from trigonometric
parallax measurements also provide a solid foundation for accurate
stellar luminosity, and thus mass, estimates with which to construct
the stellar IMF.

Through the late 1990s, most measurements of the stellar IMF based on
trigonometric parallax catalogs were restricted to the volume within
5.2~parsecs of the Sun, the completeness limit originally adopted by
van de Kamp (1971) for his compilation of nearby stars.  Jointly analyzing the luminosity functions of the 5.2~parsec sample and a magnitude limited sample of more distant stars, Kroupa, Tout \& Gilmore~(1993) demonstrated that a single IMF could be reconciled with both datasets after accounting for binarity, Galactic structure, and observational biases.  The broken power-law fit derived in these works forms the basis of the `Kroupa-type' IMF, which has been applied successfully to many galactic and extragalactic systems.

Reid \& Gizis (1997) subsequently used the compilation of nearby stars assembled by Gliese \& Jahreiss (1991) to extend the boundary of the reliably volume complete sample to a distance of 8~pc.  The most recent revision of the 8~pc sample contains 146~stars \& BDs in 103~systems, with 95\% possessing distances derived directly from trigonometric parallax measurements. 
A single power-law fit to the IMF of the 8~pc sample returns
$\Gamma$=0.2$\pm$0.3 in the 0.1--0.7~M$_{\odot}$ mass range, and
comparisons with the 5.2~pc sample limit any
significant (i.e, $> 1\sigma$) incompleteness to M$_V \sim 14$, or
equivalently, spectral types M4 or later (Reid et al.~2003).
Above this mass range, the 8~pc sample represents the present-day mass
function, and corrections must be applied based on the star formation
history of the Galactic disk to recover the IMF.  Significant effort has been invested over the past decade to extend the volume complete sample yet further, to 10~pc, and address incompleteness at low Galactic latitudes, southern declinations, and for intrinsically faint objects (e.g., Scholz et al. 2005, Henry et al. 2006, Reyle et al. 2006).  New members continue to be found
(e.g. Lepine et al. 2009, Schmidt et al. 2010), but the sample will
soon be complete enough to enable an important reduction in the uncertainties
associated with the mass function of the immediate solar neighborhood and the crucial multiplicity
correction for IMF studies containing unresolved binaries.

To achieve a more statistically precise IMF measurement than can be derived from relatively small parallax limited catalogs, many investigators have assembled larger volume complete samples of solar neighborhood field stars with less accurate photometric or spectroscopic distances.
 In his seminal study, Salpeter (1955) adopted this technique to 
measure an IMF that is well described above 1 M$_{\odot}$ by a $\Gamma$=1.35 power-law. Subsequent studies of the super-solar (M$>$1~M$_{\odot}$) field mass function have focused on the 1 -- 10~M$_{\odot}$ mass range.  The highest mass stars possess such short lifetimes, and require sufficiently complex spectroscopic
mass determinations, that it is more efficient to measure the IMF above
10~M$_{\odot}$ in OB associations (\S~\ref{sec:ob_associations}) and massive young clusters (\S~\ref{sec:ymcs}).  Studies of the super-solar IMF in the Galactic
field have measured power-law IMFs slightly steeper than, but marginally consistent with, the
canonical ``Salpeter slope". 

 As Scalo (2005) demonstrates, this apparent consistency partially reflects a
paucity of measurements of the super-solar field star IMF. We 
find only four measurements of the high-mass field star IMF
following that of Scalo (1986).   Rana \& Basu (1992), Maciel \&
Rocha-Pinto (1998), and Reid, Gizis \& Hawley (2002) all derived the
super-solar IMF by assuming specific prescriptions for the star
formation history of the Milky Way (the empirical SFHs of Soderblom,
Duncan \& Johnson~(1991), Rocha-Pinto \& Maciel~(1997) and a constant
star formation rate, respectively).  Schroder \& Pagel (2003) performed a joint
fit of the Milky Way IMF and SFH.   The high-mass IMF measurements are
all consistent with a $\Gamma$=1.65$\pm$0.2 power-law.  The only
(partial) exception is the IMF derived by Schroder \& Pagel, who
derived a slightly steeper IMF ($\Gamma$=2.1) in the 1.6 -- 4~M$_{\odot}$ regime.  In general, these measurements are in good agreement with that found in OB associations, which will be discussed in \S~\ref{sec:ob_associations}.

\begin{figure}[!h]	
\centerline{\psfig{figure=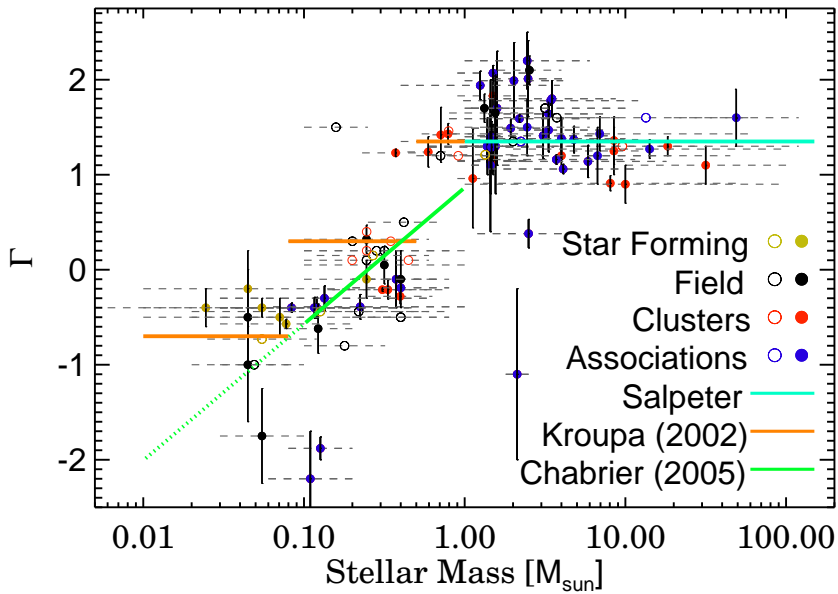,height=20pc,angle=90}}
\caption{A representation of the ``alpha plot" of Scalo~(1998) and Kroupa~(2002).  We show the derived index, $\Gamma$, of the initial mass function in clusters, nearby star forming regions, associations and the field, as a function of sampled stellar mass (points are placed in the center of log~m range used to derive each index, with the dashed lines indicating the full range of masses sampled).  The data points are from studies discussed in the text and are not meant to be a complete review of the field.  Additionally, we have added a sample of clusters compiled by Kroupa~(2002).  Open circles denote studies where no errors on the derived $\Gamma$ are given while filled circles are accompanied with the corresponding error estimate (shown as solid vertical lines).  The observed scatter in the $\Gamma$ measurements presented here is likely to be larger than in the literature as a whole, as ``outliers" are emphasized in this review. The colored solid lines represent three analytical IMFs: shown in green is the Chabrier~(2003) IMF (dashed line indicates extrapolation into the substellar regime), with a Salpeter~(1955) IMF in blue, and a Kroupa~(2002) IMF in orange (which is essentially Salpeter above 1~\msun).}
\label{fig:alpha_plot}
\end{figure}

More attention has been devoted in recent years to determining the IMF
of sub-solar (M $<$ 1 M$_{\odot}$) field stars, partially motivated by a desire to understand whether very-low mass
stars and brown dwarfs could contribute significantly to the baryonic
dark matter in galaxies (Graff \& Freese1996; Flynn et al. 1996).  Comprehensive reviews of sub-solar IMF
measurements in the Galactic field are provided by Scalo (1986),
Bessell \& Stringfellow (1993), and Chabrier (2003):  the results of a sample of 14
sub-solar field star IMF measurements are presented in Fig.~\ref{fig:alpha_plot}.    The earliest studies supplemented the
parallax limited samples of nearby (5-8~pc) stars described above with magnitude
limited samples of rarer, brighter, and more distant stars detected by
Hipparcos or in all-sky photographic surveys (Kroupa, Tout \& Gilmore
1993; Reid \& Gizis 1997; Chabrier et al. 2005).
Beginning in the mid 1990's, analyses were increasingly conducted using
data from very deep pencil-beam surveys sampling more
modest areas of the sky ($<$5~square degrees), often primarily
motivated by extragalactic science goals (Martini \& Osmer 1998; Gould
et al.~1997; Zheng et al.~2001; Schultheis et al. 2006; Robin et al.
2007; Vallinari et al. 2006).   The recent emergence of fully digital,
wide area sky surveys has enabled a new generation of mass function
analyses that leverage uniform, moderately deep photometry over a large
area of the sky to generate large, magnitude limited samples of
sub-solar field stars (Covey et al. 2008; Deacon, Nelemans \& Hambly
2008; Bochanski et al.~2010).

The underlying consensus in sub-solar field star IMF
measurements is visible in Figure 2, where the power-law slope characterizing the IMF of thin disk field
stars declines from $\Gamma=$1.7 above
1~M$_{\odot}$ to $\Gamma \sim 0$ for samples of 0.1--0.7 \msun\ field stars.
The strongest outlier to this trend is the IMF
measurement obtained by Schultheis et al. (2006), who fit optical star
counts observed in deep CFHT fields to predictions of synthetic
Galactic structure and stellar populations models (shown as an open circle at a stellar mass of 0.18\msun\ and $\Gamma=1.5$).  These models are
characterized by numerous parameters, however, of which the adopted
thin disk IMF is only one.  Schultheis et al. did not perform a robust
exploration of this multi-dimensional parameter space, and it is
possible that the star count discrepancy they interpret as a steepening
IMF could be reconciled by changing another model parameter.

Two studies have attempted to constrain the sub-solar IMF of the thick disk
(Reyle \& Robin 2001, Vallenari et al. 2006), comparing similar datasets to predictions of synthetic stellar population models of the
Milky Way.  Both studies measure a best fit thick disk IMF of $\Gamma$=-0.5 for 0.2--0.8~\msun\ stars.  No formal uncertainties are quoted in
either study,
but changes in the analysis (e.g. adopted binary prescription) induce changes
in $\Gamma$ of $\sim$0.25.  These slopes are somewhat shallower than are
observed for a similar range of masses in the thin disk.  Given the
uncertainties seen earlier in comparisons between observed star counts and
multi-parameter synthetic stellar population models, and the
difficulty that theoretical
models encounter reproducing stellar colors and luminosities in this mass and
metallicity regime, we do not consider this strong evidence for a metallicity
dependence of the IMF in the Galactic disk.

Figure~\ref{fig:alpha_plot} also includes results from four recent studies which analyzed
the IMF of sub-stellar objects in the Galactic field (Reid et al. 1999;
Allen et al. 2005; Metchev et al. 2008; Pinfield et al. 2008).  As noted in \S 1.2.1, analyses of the IMF of field brown dwarfs must model the SFH of the Galaxy
to statistically transform the distribution of cool object luminosities into a distribution of masses, inevitably introducing additional uncertainties into the measurement.   Moreover, only the sample analyzed by Allen et al. (2005)
exceeds 25 objects, limiting the strength of any conclusions that can be currently
drawn concerning the Galactic substellar IMF.  Nonetheless, the results of these Galactic field studies are consistent with a $\Gamma \sim -1.0$ substellar IMF over similar mass ranges.

\subsection{Open Clusters} 
\label{sec:open_clusters}

As bound physical objects and simple stellar populations,
open clusters are well suited to observational studies of the IMF.
Stars in open clusters possess modest (or
non-existent) age and metallicity spreads and share a common distance, obviating large uncertainties that
plague studies of the field star IMF.  Sub-stellar objects in young
open clusters
are also still relatively warm and bright, and thus relatively easy to
detect.
Unlike younger clusters still deeply embedded in their parent molecular
clouds, however, optically visible open clusters
possess modest, relatively uniform extinction, enabling greater
sensitivity to intrinsically faint objects and reducing uncertainties
associated with spatially varying reddening corrections. The one major
caveat in the study of stellar
clusters is that they dynamically evolve over time, systematically
losing their low mass stars to the field, and becoming dynamically
mass segregated (see \S~\ref{sec:segregation} and the recent review by Portegies
Zwart, McMillan, \& Gieles~2010).

Significant deviations from a Salpeter slope have been reported in
measurements of the super-solar (M$_* >$ 1~M$_{\odot}$) IMF in various
open clusters. Scalo (2005) identified the most notable discrepancies
in a review of results from the prior decade.  It should be kept in
mind, however, that these deviations represent a small minority of
cases: most analyses of large, homogeneous cluster samples (e.g.
Maciejewski \& Niedzielski 2007;  Bonatto \& Bica 2007) or
high-quality observations of individual clusters (e.g. Carraro et al.
2005; Santos, Bonatto \& Bica 2005) derive super--solar IMFs consistent
with a Salpeter slope.  In those cases where non-Salpeter slopes are
reported, the results are often attributable to low statistical
significance or systematic differences between approaches: we
summarize a few illustrative cases here. Phelps and Janes (1993)
summarized the study of several open clusters over the mass range from
about 1--8~M$_{\odot}$.  Most clusters studied possessed power--law
slopes consistent with the Salpeter value with $\Gamma = 1.4 \pm 0.13$,
but two clusters (NGC 581 and 663) exhibited signification variations
from this mean.  Incorporating proper motions as well as photometry to
select members of NGC 581, Sanner et al. (1999) confirmed a steep IMF
slope ($\Gamma = 1.8 \pm 0.2$), but the measurement uncertainties are
such that this discrepancy is only at the $\sim2 \sigma$ level.  NGC 663
has also been studied more recently: observations that cover a larger
area of the cluster indicate that the IMF is indeed consistent with
Salpeter (Pandey et al. 2005).   Sanner \& Geffert (2001) derive a
power--law slope for Stock~2 of $\Gamma = 2.01 \pm 0.4$, but comment
that the distance to the cluster, the colors of some objects, and the
dispersion in proper motions, are all problematic for this cluster.
Prisinzano et al. (2003) present a photometric study of NGC 2422,
arguing for a steeper--than--Salpeter IMF with $\Gamma = 2.07 \pm
0.08$.  A similarly steep mass function
is found for NGC~2323 ($\Gamma= 1.94$) by Kalirai et al.~(2003).
Both clusters (NGC~2422 and NGC~2323) warrant future studies
to confirm their non-Salpeter slope.

Surveying the state of the field, Sagar (2002) argue that, given the
large uncertainties in any  individual measurement, the data on
galactic open clusters are consistent with a universal IMF of roughly
Salpeter form from 1--20~M$_{\odot}$ (cf. Massey et al. 1995a/b).
Given the difficulties inherent in these studies, it would be
extremely valuable for future work in this area to address whether a
given dataset is inconsistent at the 3 $\sigma$ level with
having been drawn from the field star IMF, rather than re--deriving a
new (binning dependent) slope from the data (see \S \ref{sec:statistics}).

In recent years, many studies of open cluster IMFs have been devoted to
determining the sub-solar IMF down to sub--stellar masses.   The
Pleaides, in particular, has been the focus of significant effort.
Moraux et al. (2003) present a comprehensive study of the cluster IMF
down to 0.03~M$_{\odot}$ and find the results consistent with a
log--normal form over a broad mass range (0.03--10 M$_{\odot})$:
$m_{c} = 0.25~M_{\odot}$ with $\sigma_{logm}$ = 0.52, and a
marginally steeper ($\Gamma = 1.7$) than Salpeter slope above 1.5
M$_{\odot}$; cf. Bouvier et al. 1998).  Bihain et al. (2006) combine
photometry with proper motion information in some cases to identify
low mass members down to  0.02~M$_{\odot}$. The mass function they
derive is consistent with earlier work.  Lodieu et al. (2007a) identify Pleiades members with
wide--field infrared
photometry from the UKIDSS survey,
measuring a mass function that is consistent with Moraux et al. (2003)
and report a sub--stellar binary frequency much higher than that
characterizing older ultracool field L and T dwarfs.  Casewell et al.
(2007) present deep I and z photometry as well as proper motion
confirmation for some candidate L and T dwarf members with suggested
masses as low as 0.01 M$_{\odot}$, and derive a preliminary $\Gamma
=-0.5 \pm 0.3$ for the substellar IMF.

The Pleiades provides a key touchstone for comparison with
other open clusters.  M35 is slightly older than
the Pleiades ($\sim$ 175~Myr vs. 120~Myr), exhibits evidence for
modest mass segregation, and sports a present--day mass function
consistent with that measured in the Pleiades (Barrado y Navascues et
al. 2001).  The slightly younger $\alpha$~Per cluster (90~Myr) shows
less dynamical evolution, but possesses a mass function that is also
consistent with the Pleiades (Barrado y Navascues et al. 2002).
Jeffries et al. (2004) present a photometric study of the structure
and IMF for the even younger ($\sim30$~Myr) cluster NGC 2547.
They find evidence for (perhaps primordial, although see
\S~\ref{sec:segregation}) mass segregation above 3~M$_{\odot}$, an IMF
similar to the Pleiades between 0.07-0.7~M$_{\odot}$ and tentatively
suggest a dearth of sub--stellar objects in the cluster.  The young
(50--100~Myr) cluster IC 4665 was studied by de Wit et al. (2006), who
also find a mass function similar to that of the Pleiades down to 0.04
M$_{\odot}$.  Moraux et al. (2007a) present the IMF for the 100--150
Myr old cluster Blanco I, again finding a log--normal to be a good fit
between 0.03--3.0 M$_{\odot}$ with $m_{c} = 0.36 \pm 0.07
M_{\odot}$ and $\sigma_{logm} = 0.58 \pm 0.06$.

Disentangling IMF variations from dynamical evolution becomes
more difficult for older open clusters.  Bouvier et al. (2008)
studied the $\sim$625 Myr Hyades cluster, which shows signs of
significant dynamical evolution, but is consistent with the
Pleiades IMF above about 1.0~M$_{\odot}$.  Boudreault et al.
(2009) recently obtained a mass function for the similarly aged (590 Myr)
Praesepe cluster,
however, finding no deficiency in low-mass members as expected from dynamical
evolution.  This wealth of low-mass members is also present in the Praesepe IMF constructed by Kraus \& Hillenbrand (2007).  Neither study definitively confirmed each star's membership with
spectroscopic
observations, but Kraus \& Hillenbrand did incorporate proper
motions into their membership analysis, obtaining a $\Gamma=0.4\pm0.2$ power-law
slope over
the  mass range $0.12-1.0~M_{\odot}$, consistent with the field star IMF.
Boudreault et al.,
by contrast, find a steeper $\Gamma=0.8\pm0.1$ IMF over the narrower
0.15--0.5 M$_{\odot}$
mass range.  Disagreement between the Praesepe and Hyades present day mass functions
could arise from variations in the cluster's IMFs, or from differences in the
dynamical evolution each cluster has undergone.  Confirming the differences
between the present day mass functions of these benchmark clusters, and developing a better
understanding of their origin, is a promising area for future work.

\begin{figure}[!h]	
\centerline{\psfig{figure=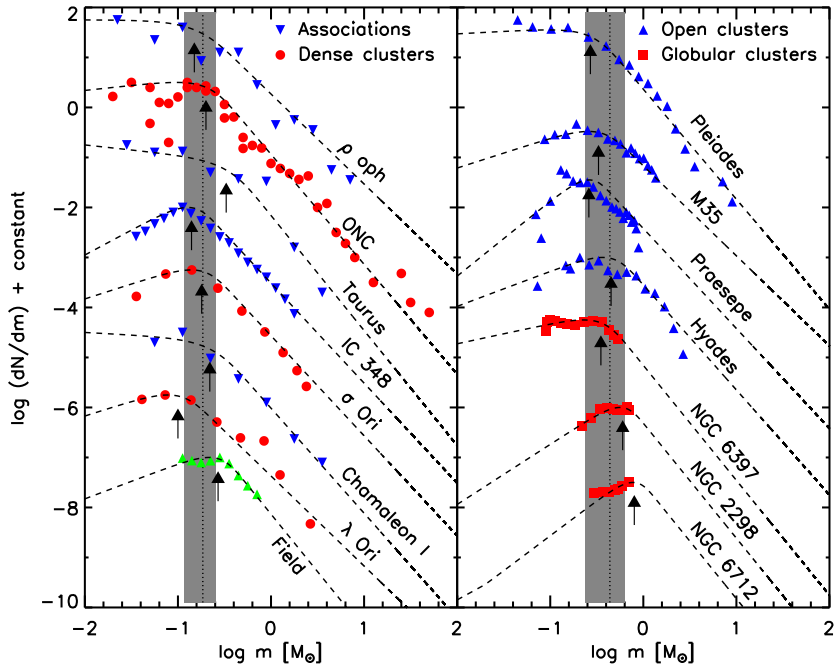,height=22pc}}
\caption{The derived present day mass function of a sample of young star-forming regions (\S~\ref{sec:nearby_young_clusters}), open clusters spanning a large age range (\S~\ref{sec:open_clusters}), and old globular clusters (\S~\ref{sec:globulars}) from the compilation of de Marchi, Parsesce \& Portegies Zwart~(2010).  Additionally, we show the inferred field star IMF (\S~\ref{sec:field}).  The dashed lines represent ``tapered power-law" fits to the data (Eqn.~\ref{eqn:tapered}). The arrows show the characteristic mass of each fit ($m_p$), the dotted line indicates the mean characteristic mass of the clusters in each panel, and the shaded region shows the standard deviation of the characteristic masses in that panel (the field star IMF is not included in the calculation of the mean/standard deviation).  The observations are consistent with a single underlying IMF, although the scatter at and below the stellar/sub-stellar boundary clearly calls for further study.  The shift of the globular clusters characteristic mass to higher masses is expected from considerations of dynamical evolution.}
\label{fig:guido}
\end{figure}

In summary, nearly all open clusters exhibit
present day mass functions that resemble (and many are formally
consistent with) a dynamically evolved Kroupa/Chabrier type IMF (see
Figs~\ref{fig:alpha_plot} and \ref{fig:guido}).  In a recent study, de Marchi, Paresce, and Portegies Zwart~(2010)
compile a list of open clusters (along with nearby star-forming
regions and globular clusters) from the literature and fit these mass
functions with tapered power laws (Eq.~\ref{eqn:tapered}).  For the
denser clusters, where mass segregation is likely to have occurred,
the authors determine the mass function for the full (global) cluster,
or if not possible, at the half light radius which is not expected to
be strongly affected by mass segregation.  They find
that all young clusters and star-forming regions are well fit by
$\alpha=2.1\pm0.2$ (consistent with the Salpeter value), $\beta
=2.4\pm0.4$, and $m_p = 0.23\pm0.1~\msun$.  A selection of their full
sample is shown in Fig.~\ref{fig:guido}.  Some of the dynamically older clusters
in their sample (e.g. M35, the Pleiades, and globular clusters) have
systematically higher $m_p$ values, however this shift likely
represents dynamical evolution where the lowest mass stars ``evaporate" (Baumgardt et al.~2008; Kruijssen~2009).  Hence
there is an expected, and observed, correlation of $m_p$ with the
cluster relaxation time (de Marchi
et al.~2010).

\subsection{Young Clusters and Associations} 
\label{sec:nearby_young_clusters}

\subsubsection{Primordial and Dynamical Mass Segregation}
\label{sec:segregation}

An additional complication in IMF studies comes from the spatial distribution of stars within a cluster or association.  The most massive stars in large, young clusters are often located in a cluster's innermost regions. This phenomenon is known as ``mass segregation" and there are two proposed causes.  The first is that mass segregation is primordial, with massive stars forming preferentially at the center of the cluster potential.  Primordial mass segregation requires that high density regions within a star-forming cloud form a larger proportion of massive stars than lower density regions.  Theoretical explanations for primordial mass segregation range from coagulation models where massive stars are formed through coalescence, to competitive accretion models where massive stars form in the center of a potential well, to high mass stars forming in high pressure turbulent cores with high infall rates (see Zinnecker \& Yorke~2007 for a recent review). Primordial mass segregation would represent a clear violation of the hypothesis of a spatially and temporally invariant IMF, even if only over relatively small spatial scales.

The second explanation is that mass segregation is simply the result of a cluster's dynamical evolution, where more massive stars sink to the center of the cluster in about a relaxation time due to energy equipartition (e.g.~Binney \& Tremaine~1987).  Mass segregation in young clusters is often regarded as evidence for primordial mass segregation since their ages are smaller than their current relaxation timescale, and thus there has not been sufficient time for dynamical mass segregation to take hold.    The Orion Nebula Cluster provides a useful example:  it has been known for nearly a century that the cluster's most massive stars (the Trapezium stars) sit in the cluster's center, and Bonnell \& Davies (1998; see also Hillenbrand \& Hartmann 1998) concluded this mass segregation was primordial, a relic of the cluster's initial state. Several recent theoretical studies, however, suggest that the timescale for dynamical mass segregation may be shorter than previously thought.  Many young clusters are currently expanding (Mackey \& Gilmore 2003; Bastian et al.~2008), and possibly formed through the merger of several smaller sub-clusters.   Both of these effects suggest that stars in clusters may have inhabited smaller, denser environments in the past, with correspondingly shorter relaxation timescales and faster dynamically evolution into a mass segregated state (McMillan et al.~2007; Allison et al.~2009; Moeckel \& Bonnell~2009).  

The literature abounds with reports of mass segregation, but confident detections are difficult.  The challenge is that observational selection effects (e.g. crowding, brightening completeness limit near the core) produce the same observational signatures as genuine mass segregation (Ascenso et al.~2009).   Allison et al.~(2009) and Moeckel
\& Bonnell~(2009) have introduced a new method to detect and quantify mass segregation, based on the average distance between the massive stars, relative to a randomly drawn sample of lower mass stars.  Their methods appear to overcome many of the shortcomings of previous methods (i.e. it is not biased by the presence of substructure) and the authors have begun applying their method systematically to a sample of young clusters, in particular the Orion Nebula Cluster.  Future kinematic studies will be extremely valuable for assessing the dynamical  state of young clusters and  will be able to attribute differences in present-day spatial distributions as a function of stellar mass to either formation processes or the effects of dynamical evolution (e.g. Proszkow \& Adams 2009). 

Due to the observational effects of present day mass segregation, it is important to measure the global mass function of a cluster, not just the inner regions.   In some extreme cases, dynamical evolution may force a significant number of low mass stars out of the cluster altogether, hence the dynamical state of a cluster is important to note when carrying out mass function studies of very dense stellar systems.  This issue is dealt with in detail in Portegies-Zwart, McMillan \& Gieles~(2010).

\subsubsection{Young and embedded clusters}
The vast majority of stars are thought to be born in unbound aggregates that
dissipate as they disperse their
parental molecular cloud (Lada \& Lada 2003).  It is
therefore vital to determine if all star forming events
sample a single, universal IMF, or if the field star IMF merely
represents the average distribution produced from IMFs unique
to each region. Deep measurements of the IMF within embedded clusters may even detect a minimum mass below which no cluster
members exist; this limit would presumably correspond to mass of the
smallest clumps in a cloud which can undergo thermal fragmentation
(Low \& Lynden-Bell~1976; Rees~1976; Spitzer 1978; Boss 2001; Larson 2005; Whitworth et al.~2007; see
extended discussion in \S 4). The minimum mass for fragmentation is the final product of the convolution of a number of physical
parameters (e.g. density, temperature, pressure, metallicity, magnetic
field strength, turbulent velocity spectrum, etc.) which could vary
from cloud to cloud, and whose influence we could hope to infer from
measurements of the stellar IMFs in differing cloud
environments\footnote{We note, however,
that inferring a cloud core's initial conditions from its post-star
formation properties is a bit like sifting through the wreckage of a
car that was hit by a train to determine what radio station the driver
was listening to.
}.

The unique opportunities that young clusters provide for IMF studies are,
however, associated with significant challenges.  
A number of complications
must be addressed
to empirically measure a young
star's temperature and bolometric luminosity (and thus infer the mass and age through comparison with theoretical models): non-photospheric
emission in the UV/blue (from material accreting onto the star) and
infrared (from the circumstellar disk), significant extinction due to dust
along the line of sight to the star, and surface gravity effects that
shift the conversion
from spectral type to temperature.  Differences in input physics
adopted in the theoretical models required to infer each star's mass and age
introduce additional uncertainties: systematic offsets between the
masses and ages
predicted by different model grids can be as large as a factor of two,
and the unknown accretion history of a given YSO can introduce
additional random/stochastic errors (e.g.~Hartmann, Cassen, \&
Kenyon~1997; Baraffe, Chabrier \& Gallardo 2009).
Comparing model predictions for pre-main sequence stars with robust
dynamical mass measurements indicates that mass errors at the 30--50\%
level are typical (Hillenbrand \& White~2004), and errors in more
poorly characterized
systems are likely larger still.

Despite these challenges, much work was invested in the 1990s to search for variations in the
IMF as a function of initial conditions.  Many infrared photometric surveys
of several embedded clusters were undertaken, critically supported by optical
and infrared spectroscopy to estimate the cluster's age and adopt an
appropriate mass--luminosity relationship.  The wide range of extinctions
present within a cluster, ranging from $A_V < 2.0^m$ for the cloud's surface
population to $A_V > 30^m$ for stars buried in the cloud's core, creates an
observational bias whereby the cluster's most luminous members (i.e.
the youngest and/or
highest mass stars) can be detected to greater depth within the cloud.
 These bright
stars are therefore over-represented in a magnitude limited sample,
necessitating the
construction of an extinction-limited sub-set of the cluster's members
to produce an
unbiased measure of the cloud's IMF (Meyer et al.~2000).

These techniques have been applied extensively to the Orion Nebula
Cluster, a valuable young cluster
for star formation studies due to its proximity and richness (see
review by Muench et al. 2008).
Numerous studies of the IMF of the
Orion Nebular Cluster have been made over the past decade 
(Hillenbrand 1997, Hillenbrand \& Carpenter 2000, Luhman et al. 2000,
Muench et al. 2000, 2002, Slesnick et al. 2004, Lucas et al. 2005,
Weights et al. 2009).
These independent IMF determinations are strikingly consistent,
deriving results that
strongly resemble a Kroupa/Chabrier-type IMF: a Salpeter-like slope in
the super-solar
regime that breaks below 1 M$_{\odot}$ to a broad peak (in logarithmic space)
at 0.2-0.3 M$_{\odot}$.  The Orion IMF is well sampled across a wide range of
stellar masses, providing a useful benchmark for comparison with the IMF shapes
measured in different star--forming regions.  These comparisons require yet
more care, however, to avoid systematic errors arising from
gross differences in the underlying models and observations, and more
subtle effects such as the identification and treatment of unresolved multiple
systems.  As an example, Meyer et al.~(2000) perform a K--S test between the
stellar mass distributions measured in the Pleaides (Bouvier et al 1998) and
the Orion Trapezium cluster (Hillenbrand 1997). The results suggest a very
small probability that they share the same parent population, but this
most likely reflects systematic differences in the way the masses were derived
rather than a true
astrophysical difference.   Summarizing results from several studies,
Meyer et al.
(2000) were able to demonstrate that the IMFs between 0.1--10
M$_{\odot}$ in young
clusters were: a) consistent with each other; b) consistent with having
been drawn from the field star IMF; and c) NOT consistent with a Salpeter
power--law slope IMF extended down to the hydrogen--burning limit.

The Taurus dark cloud, one of the nearest sites
of active star formation suffering relatively low extinction, has been the target of intense observational
scrutiny for decades.  Hundreds of young stellar
objects have been identified as candidate association members, located
throughout the boundary of the molecular gas and beyond (Kenyon, Gomez
\& Whitney~2008).
Yet the completeness of these surveys as a function of stellar
mass remains a controversial topic today.  The distribution of spectral types
for known Taurus members has a significant peak at K7--M0, although until
recent years this was assumed to be largely due to selection effects.  
The size of the Taurus dark cloud complex ($>$ 20 square degrees) makes it
 difficult
to survey the whole cloud with the sensitivity required to detect
objects near the
hydrogen--burning limit, but recent surveys have covered larger areas
to increasingly
impressive photometric depths.
Briceno et al. (2002) performed the first statistical test of the apparent
discrepancy in the distribution of spectral types in Taurus.  They
found a 0.1 \% chance
that the distribution of objects in the Taurus dark cloud with masses
between 0.02--1.0 M$_{\odot}$ is consistent with that found
in the Trapezium: the Trapezium appeared to have relatively more brown dwarfs
than Taurus.  Recent surveys have identified more brown dwarfs in
Taurus (e.g. Luhman 2004, Guieu et al. 2006),
which diminished this apparent difference, but the observed IMF in Taurus remained anomalous
with respect to the field and other star forming regions.

Most recently, Luhman et al. (2009) analyzed
Spitzer observations and x--ray data obtained
from the XMM satellite (XEST - XMM-Newton Extended Survey of
Taurus; G{\"u}del et al.~2007). Follow--up spectroscopy
of candidate members detected in the XEST survey and
comparisons to catalogs of previously known members suggests that
the fields covered should be complete down to 0.02 M$_{\odot}$ for
the more evolved Class III young stellar objects, but
that incompleteness  may affect the census of Class I and late type (i.e., M5 or later)
Class II objects. Luhman et al. (2009) perform a K--S test to compare the IMF they
measure from the XEST fields to those previously measured using similar
methods in IC 348 (Luhman et al. 2003) and Chameleon I (Luhman 2007):
they find a 0.04\% chance that the Taurus IMF could be drawn from the same
parent distribution as the IC 348 and Cha I IMFs.  The best data
available therefore
indicate a significant difference between the Taurus IMF and that found
in other young star forming regions.

Infrared surveys of Taurus continue to find new members, however, which may yet
again bring the IMF of Taurus into closer agreement with other young
star forming regions.  Rebull et al.~(2010) obtained
spectroscopic follow-up of candidate young stars identified in the 
Spitzer Taurus Legacy Survey, confirming 37 new Taurus
members, 16 of which
appear to be new identifications that were not present at the time of
the Luhman et al. analysis.
Half of these possess spectral types of M5 or later,
highlighting the difficulty of identifying a complete census of the
lowest mass members of
Taurus. Nonetheless, the peak of K7/M0 members in Taurus is
sufficiently strong that
many more new late-type members would need to be identified to reconcile the discrepancy.
 If no such sample is
forthcoming, it will represent a significant break--through in the search
for IMF variations as
a function of initial conditions.  The next task will be understanding which
of Taurus' many unusual properties (low total mass, low stellar
density aggregates, apparent coordination of star formation
across the cloud) is responsible for its anomalous IMF.

Taurus represents the only star forming region where significant
IMF differences have been reported in the stellar regime, but unusual substellar IMFs have been reported in
various regions.  A review of sub--stellar objects in young
star--forming regions can be found in Luhman et al. (2007).  We focus here only
on the reports of IMF variations.
 Lyo et al. (2006) report that the IMF of $\eta$ Cha is
deficient at sub--stellar masses.  Moraux et al. (2007b) explored
whether this apparent deficiency could be due to dynamical evolution,
but Luhman et al. (2009) find that the deficiency is not statistically
significant.
An excess, rather than a dearth, of brown dwarfs has also been reported
in $\sigma$ Ori (B\'{e}jar et al.~2001; Caballero et al.~2007).  Sherry
et al. (2004) report that the stellar portion of $\sigma$ Ori's IMF is consistent
with the field star IMF, but an analysis of UKIDSS observations
of the cluster by Lodieu et al. (2009) indicates preliminary evidence for
mass segregation and an IMF that rises smoothly across the stellar
sub-stellar boundary.  An excess of brown dwarfs above what is seen
in the field has also been reported in Upper Sco.  Lodieu et al. (2006)
identified candidate members using UKIDSS photometry (with subsequent
spectroscopic confirmation for most candidates; Lodieu et al. 2007b) and
measured an $\Gamma= -0.4\pm0.1$ IMF from 0.01--0.3~M$_{\odot}$.
Slesnick et al. (2008) also find more brown dwarfs in Upper Sco  than
reported for the field or other star-forming regions, with the IMF
peaking at 0.1~\msun\ compared to 0.2-0.3~\msun.  These differences in
the substellar IMF are tantalizing, but their statistical significance
has not yet been determined via rigorous tests for consistency with the field star IMF.

Statistical tests that have been applied to the substellar IMF
have resorted to comparisons of the ratio of stars to substellar
objects.  This ratio is essentially an extremely coarse sampling
of the IMF that reduces the importance of random errors in
inferred brown dwarf masses.  Star-brown
dwarf ratios have been calculated for several nearby star forming
regions, including Taurus (Briceno et al. 2002, Luhman et al. 2003a,
Luhman 2004), IC 348 (Luhman et al. 2003b), Orion (Luhman et al. 2000,
Hillenbrand \& Carpenter 2000, Muench et al. 2002, Slesnick et al.
2004), and Chameleon I (Luhman 2007).  These studies indicate apparent 
cloud-to-cloud variations in the star-brown dwarf ratio at the factor
of two level, but do not provide strong evidence for IMF variations
given the uncertainties involved.
Merging their own data with
that of Wilking et al. (2004), Greissl et al. (2007) attempted to construct the
stellar/sub-stellar ratio in NGC 1333.  While each study had measured
an IMF consistent with that of the field over a different mass range,
Greissl et al. were unable to reconcile the combined stellar/sub-stellar ratio with that
found for the field IMF.   Scholz et al.
(2009) have since
measured a ratio of stars to sub--stellar objects consistent with that
of the field,
but they suggest the presence of a low-mass cutoff below 0.02~M$_{\odot}$.
Andersen et al. (2008) calculated the ratios of stars to sub--stellar objects
from studies of seven star forming regions (Mon R2: Andersen et al. 2006;  the
Pleaides: Moraux et al. 2003; NGC 2024: Levine et al. 2006; and
Taurus, the Orion Nebular Cluster, Chameleon I, and IC 348: see above
for references for these latter clusters)
in the mass range 0.03--1.0~M$_{\odot}$.
They found that these ratios were: a) consistent with a single underlying IMF;
b) that the ensemble data are inconsistent with an IMF that rises into
the sub--stellar regime (i.e. requires $\Gamma < -1.0$ );
and c) the star-brown dwarf ratios are consistent with extrapolation of the log--normal
form of the stellar IMF into the sub--stellar regime.

To summarize, most star forming regions appear to have stellar 
IMFs consistent with that seen in Orion and the Galactic field.
The most significant differences found to date are those
between Taurus and other well--studied regions, and,
perhaps between Upper Sco and the Trapezium and the Pleiades.
Below the hydrogen burning limit, most star forming regions appear to possess sub-stellar 
IMFs with $\Gamma <\sim -0.5$.

\subsubsection{OB Associations}
\label{sec:ob_associations}

While OB associations are not as dense as the clusters which will be discussed in the next section, their capacity to form large numbers of high mass stars makes them very important sites of star-formation in terms of galactic chemical enrichment and ISM energetics.  Optically--visible OB associations are attractive regions to study the IMF above 1.0 M$_{\odot}$ before they disperse into the field.  
Photometric surveys covering the UBV bands can provide initial estimates of
extinction, temperature, and luminosity, with follow-up spectroscopy required
to distinguish between stars evolving towards, or away from, the main sequence
in the HR diagram. In a comprehensive study of OB associations in the Milky Way as 
well as other Local Group galaxies, Massey et al.  (1995a/b) found that the IMF 
for high-mass stars appears to follow a standard $\Gamma$=1.35 Salpeter slope,
and that it is not sensitive to metallicity.  They do note, however, that isolated OB 
stars appear to have a power--law mass distribution with a steeper slope 
(see Massey~2003).  More recent studies of 
massive star--forming regions within a few kpc of the Sun also find power--law IMFs in the high-mass
regime.  Slesnick et al. (2002) find $\Gamma = 1.3 \pm 0.2$ for the double cluster h and $\chi$ Persei,
with similar results reported for the massive star forming regions  W49 ($\Gamma=1.6$, $20 < \msun < 120$; Homeier \& Alves~2005), NGC~3576 ($\Gamma=1.6$, $3 < \msun < 60$; Figuer{\^e}do et al.~2002), and a host of other H{\sc ii} regions in the Galaxy.  

Finally, we point the reader to the review of Massey~(2003; see also Zinnecker \& Yorke~2007) who summarized the measured value of $\Gamma$ for massive stars in a variety of Galactic OB associations and clusters that were all analyzed in a consistent fashion.  While there is variance in the measured values, the majority of the observations (and the mean) are consistent with the Salpeter value.  We also note that some of these values have changed considerably due to new deeper imaging (also the application of new stellar isochrones), e.g. Cyg~OB2 has $\Gamma=0.9$ in the Massey et al. (1995a,b) compilation, $\Gamma=1.6$ in the study by Kn{\"o}dlseder~(2000) and $\Gamma=1.27$ in the most recent study of Wright \& Drake~(2010).

\subsection{Extreme Star Formation in the Milky Way}

If the IMF does vary with environment, extreme star formation sites may have mass functions which deviate significantly from what we see locally.  Some such ``extreme sites" are found in the Galaxy, namely in "starburst clusters" which have stellar densities orders of magnitude larger than nearby clusters.  Additionally, the Galactic center is equally exotic, as the dynamics are dominated by a super massive black hole.  Studies have reported strange, non-canonical, IMFs for both environments, and in this section we investigate these claims in turn.  

\subsubsection{Starburst Clusters}
\label{sec:ymcs}

The discovery of young Milky Way clusters with masses and densities comparable to or larger than Galactic globular clusters has prompted a major paradigm shift in our understanding of cluster formation.  
These young clusters are analogous to the extragalactic ``starburst clusters" that will be discussed in \S~\ref{sec:ssc}, but due to their proximity, can be studied in much greater detail.  Most of the extreme clusters in our Galaxy reside at distances of 3-8~kpc from the Sun, so we cannot study their mass functions in comparable detail to clusters in the solar neighborhood.  Nevertheless, they provide crucial laboratories to examine if the IMF varies in extreme star forming environments.

The young ($2-4$~Myr) Arches cluster is a prime example.  Located $\sim25$~pc away from the Galactic center in projection, the Arches is one of the most massive and densest ($\rho \simeq 2\times10^5$~\msun~pc$^{-3}$) clusters in the Galaxy, and it has been the focus of numerous studies to determine its mass function and any variance with cluster radius.  In the first such study, Figer et al.~(1999) used NICMOS on HST to measure the mass function above $\sim6$~\msun, finding a mass distribution that was significantly flatter ($\Gamma$=0.65) than the Salpeter value.   Using adaptive optics (AO) assisted near-IR imaging on Gemini-North, Stolte et al. (2002) subsequently found a slightly steeper slope ($\Gamma$=0.8 over the same mass range), but still significantly flatter than the Salpeter law.  Stolte et al. (2002) also reported a turnover in the mass function at $\sim6$~\msun, implying that the characteristic turnover mass is $\sim20$ times higher near the Galactic center than it is locally.  Kim et al.~(2006) observed the Arches with the AO system on Keck, revealing a yet slightly steeper IMF ($\Gamma$=0.9).  These authors also found that the mass function continues down to $\sim1.3$~\msun\ as a single power-law, and showed that the reported turnover at $\sim6$~\msun\ was either a local bump, or simply an artificial excess of stars due to the conversion of stellar luminosities to mass.  Most recently, Espinoza, Selman \& Melnick (2009) used AO on the VLT to measure the Arches' global mass function index above $\sim$10~$\msun$, paying particular attention to differential extinction, obtained a present day IMF index of $\Gamma$=1.1$\pm$0.2, marginally consistent with the Salpeter value.  

All the studies of the Arches described above report strong radial gradients in the index of the mass function, indicating the presence of a large amount of mass segregation.  Kim et al. (2006) calculated that, given the age of the cluster, dynamical mass segregation could flatten the observed IMF index by $\sim$0.1-0.2.   However, if the cluster formed via the hierarchical merging of sub-clumps (e.g. McMillan et al.~2007; Allison et al.~2009) then the current IMF may in fact tell us more about the dynamical history of the cluster than the IMF.

Given the distance to the Arches cluster, observations have so far not been able to detect individual sub-solar mass stars.  In a complimentary approach to resolved studies of individual stars, Wang, Dong \& Lang~(2006) have attempted to constrain the low mass IMF by comparing the diffuse X-ray emission of the Arches with the amount expected by extrapolating the X-ray flux of the Orion Nebula Cluster.  Wang, Dong \& Lang conclude that the Arches likely has an order of magnitude fewer low-mass stars than predicted by a Miller \& Scalo (1979) IMF, normalized to match the number of M $> 60 \msun$ stars identified in the Arches by Stolte et al. (2002).  The log-normal form of the Miller \& Scalo IMF, however, produces an extremely steep ($\Gamma$=2.3) IMF for masses $>$ 60 $\msun$, leading to an extremely large low-mass population given a fixed number of high-mass stars.  Indeed, Wang, Dong \& Lang demonstrate that the Arches diffuse X-ray emission is consistent with that expected from the low-mass population implied by a simple extrapolation of the Stolte et al. (2002) $\Gamma$=0.9 IMF, even without invoking a sub-solar flattening of the IMF.  Recently, much more sensitive X-ray images of the Arches have been obtained (Hong et al.~2009); a similar analysis of the diffuse X-ray emission in these images may provide an improved understanding of the Arches' sub-solar IMF.

The most massive known young cluster in the Galaxy, Westerlund 1 ($4-7$~Myr), is slightly older than the Arches cluster ($2-4$~Myr) and is one of the closest young massive clusters ($3-5$~kpc from the Sun).  Brandner et al.~(2008) derived the mass function within the inner 1.5~pc of the cluster for stars with masses between 3.4 and 27~\msun, and found $\Gamma=1.3$, consistent with the Salpeter value.  They also report a significant flattening in the IMF in the inner parts of the cluster and a steepening in the outer parts, showing that this cluster also has significant mass segregation.

NGC~3603 is another young ($2-3$~Myr), dense and massive cluster in the Galaxy, located $\sim6$~kpc from the Sun (Stolte et al.~2004).  Nurnberger \& Petr-Gotzens~(2002) used the K-band luminosity function of stars to constrain the IMF and found that it was consistent with the Salpeter distribution for stars with masses above 0.5~\msun.  Using high-resolution VRI HST data, Sung \& Bessell~(2004) derived $\Gamma=0.9$ for the inner 20" ($\sim0.6$~pc) of the cluster over the mass range of 1-100~\msun.  Stolte et al.~(2006) confirmed this value using deep near-IR VLT imaging.  Harayama, Eisenhauer \& Martins~(2008) used AO near-IR imaging and report a flatter mass function, $\Gamma=0.74$ from $0.4-20$~\msun.   As in the other starburst clusters, all the above studies found evidence for a radially varying mass function, i.e. evidence for mass segregation.
  
 Many of the clusters discussed in this section (see also \S~\ref{sec:ssc}) contain large numbers of high mass stars which are expected to significantly affect their surroundings.  For example, these clusters may be expected to lack low mass stars, as the formation timescales of low mass stars are longer than the lifetime of the most massive stars, the presence of large numbers of ionizing stars could disrupt the natal cloud and terminate further low mass star formation (c.f. McKee \& Ostriker~2007).    Hence the shape of the low mass end ($<1\msun$) of the IMF in these clusters, which will be accessible with the next generation of telescopes, will be extremely important in understanding the effects of feedback.  Additionally, the presence of feedback can have large implications as to how a given IMF is sampled, i.e. in what order the stars are formed, a point that we will return to in \S~\ref{sec:igimf}.
  
\subsubsection{Galactic center}

The central region of the Galaxy is a rather unique place as the
dynamics are heavily influenced by the presence of a massive black hole,
M$_{\rm BH} \sim 3 \times 10^6 M_{\odot}$ (e.g. Ghez et al.~2003; Sch{\"o}del et al.~2003).  The
tidal forces due the massive black hole should render the gravitational
collapse from a cold molecular cloud virtually impossible (Morris~1993).
Thus it is surprising that a large population
of young stars exists within the central parsec (e.g. Krabbe et al.~1995).
The origin of these stars is actively debated: did they form
in situ or outside the central parsec, later arriving at
their current positions dynamically?  There is also a nuclear cluster
comprised of stars that formed at a near constant rate which is surrounded
by  two warped disks of young stars with ages of $6\pm1$~Myr
(Paumard et al.~2006; Bartko et al.~2009).  Outside the region of these disks is again
a normal stellar population.  How the disks formed (in situ or through the
spiralling in of a dissolving star cluster due to dynamical friction) is
still under investigation. These stellar populations in the Galactic
center represent an extreme case of star formation and as such their mass
functions have been under intense scrutiny.

Using adaptive optics assisted integral field observations, Paumard et al.~(2006) identified 41~OB supergiant, giant and main sequence stars.  From this sample they generated the K-band luminosity function in two ways.  The first was a sample of all stars, where late type stars (those not thought to be associated with the inner parsec) have been removed.  The second was a selection based on orbit (i.e. associated with one of the disks in the inner region) and projected distance from the Galactic center.  The derived K-bad luminosity function was substantially flatter in the first method, indicating that the results were  dependent on how the sources were selected.  They then constructed the expected luminosity function based on a Salpeter IMF, normalized to pass through the observations at a given magnitude.  The predicted luminosity function for a Salpeter IMF is steeper than the observations and the authors suggest that  a much flatter IMF ($\Gamma = -0.15$ compared to $\Gamma_{\rm Salpeter} = 1.35$) is needed to fit the observations.  
However, no statistical comparison was carried out between the observations and the expected K-band luminosity function of a Salpeter IMF.   The authors also note that a large section of the observations ($m_K < 11$) is dominated by very bright evolved stars whose evolution is quite uncertain.  The adoption of a single K-band extinction estimate for the entire region may contribute additional uncertainty to the derived IMF, as subsequent work identified a range of extinctions ($\Delta A_K \sim 1$) in this cluster (Maness et al.~2007).

Bartko et al.~(2010) present an updated adaptive optics integral field study of the Galactic center and split their sample into three subsamples based on projected radius and dynamics.  In the very inner region ($<0.8"$, $<0.03$~pc) the observed KLF was consistent with that expected from a Salpeter IMF, and so were the outer regions ($12" < R <  25"$).  However, the region between these two, which includes the nuclear stellar disks ($0.8" < R < 12"$) was much flatter.  These disks represent a particularly extreme environment of star formation and further detailed studies of them and their origin would be very insightful.

Extremely flat (i.e. top-heavy) IMFs also produce large amounts of stellar mass black holes, due to the over-abundance of high mass stars.  These dark objects will have a large impact on the dynamics of the Galactic center and it appears their presence in the nuclear stellar cluster is inconsistent with observations of stellar motions (Loeckmann, Baumgardt, \& Kroupa~2010).  Additionally, Loeckmann, Baumgardt, \& Kroupa report that the Galactic center is consistent with a canonical (e.g. Kroupa/Chabrier-type) IMF and a constant star formation history by modelling the mass-to-light ratio and total stellar mass. 

Nayakshin \& Sunyaev~(2005) probed the mass function of the inner few parsecs of the Galaxy, using the ratio of high mass stars to X-ray flux as a proxy for the IMF.  Nayakshin \& Sunyaev found that the Galactic center X-ray flux was significantly lower than expected from scaling Orion's X-ray flux by the same factor required to reconcile the number of high mass stars in each region, indicating a deficit of low mass stars in the Galactic center.  These observations motivate follow-up by the next generation of AO assisted telescopes, which should be able to directly measure the Galactic center IMF into the sub-solar mass regime. 

\section{The Local Universe}

We now leave the disk of the Milky Way behind and consider extragalactic investigations of the IMF.  We begin with our nearest galactic neighbors, the Magellanic clouds and other Local Group galaxies.  For these, we are still in a regime where individual stars can be resolved with high resolution imaging, although the distance to these objects precludes us from probing substantially into the sub-solar mass regime.  We then consider the integrated properties of parts of these galaxies and those further afield, namely their associations and stellar clusters.  As we move further from the Galaxy we begin to deal only with the integrated properties of full galaxies and we consider reports that the integrated galactic initial mass function (i.e. the sum of all the IMFs of individual star forming regions within a galaxy) may be different than the IMF of a single region.  We pay particular attention to recent studies that compare the expected H$\alpha$ and UV luminosities of galaxies and individual regions in order to place constraints on the IMF.  We conclude this section with an analysis of studies on the dynamical properties and chemical evolution of galaxies, and what they suggest regarding IMF variations.

\subsection{Nearby Galaxies}

\subsubsection{The Large and Small Magellanic Clouds}
\label{sec:lmc}

Due to their proximity, the Magellanic Clouds offer an excellent opportunity to study the IMF in lower metallicity environments.  Massive star-forming regions (e.g. R136) in the LMC provide additional opportunities to test for IMF variations in quiescent and extreme environments, while the SMC allows us to probe the IMF outside of disk galaxies.

The 30~Doradus region, and the central cluster R136, is unique in the LMC, having a mass which rivals the young Westerlund~1 in the Galaxy with an age of $\sim3$~Myr.  Due to its prominence, 30~Dor has been the subject of many IMF studies, with often contradictory results,  highlighting the difficulty in determining the IMFs of even relatively nearby systems.  Brandl et al.~(1996) used adaptive optics imaging to derive a global mass function for R136 with an index of $\Gamma=1.6$ for stars more massive than $\sim$5 $\msun$, but with a strong radial gradient that is consistent with dynamical mass segregation.   Massey \& Hunter~(1998) revisited R136 using HST photometry and spectroscopy, and found the IMF index to be $\Gamma=1.4$ from $2.8-120$\msun.      Sirianni et al.~(2000) used deeper HST V and I photometry to push to lower masses and reported a flattening in the IMF at $\sim2$~\msun.  Using NICMOS data, Andersen et al.~(2009) show that the low mass end is consistent with the predictions of a Chabrier type IMF and explore the effects of differential reddening on previous work.  In addition, Selman \& Melnick
(2005) measured a $\Gamma=1.35$ from 7--40 \msun in the 30~Dor region as a whole (excluding R136).

Kerber \& Santiago~(2006) similarly analyzed five stellar clusters in the LMC and, while they found evidence for mass segregation in all, the global present day mass functions for all five were consistent with a Salpeter distribution from 0.9 $\leq m/\msun\ \leq 2.5$.  The clusters range in age from 10~Myr to 1.75~Gyr, and span a factor of 10 in central densities, with the lowest density cluster (NGC 1818) being $\sim$3 orders of magnitude less dense than R136, and yet still well fit with a Salpeter slope above 1 M$_{\odot}$.   Hunter et al.~(1997) and de Grijs et al.~(2002) also found an index for the IMF of NGC~1815 consistent with a Salpeter value from $1-10$\msun.  Additionally, Da Rio et al.~(2009) found that the IMF in the young star-forming region LH~95 in the LMC was consistent with a Kroupa/Chabrier type IMF into the subsolar regime ($>0.4$~\msun).

The SMC has a lower star formation rate than the Galaxy or LMC, and it is not currently producing clusters as massive as R136.  Its irregular morphology and sub-LMC metallicity make it an interesting object in which to search for IMF variations.  Recent work using deep imaging with ACS and WFPC2 on HST have measured mass functions in SMC clusters which appear to be indistinguishable from the classical Salpeter value.  Examples include NGC~346 (Sabbi et al. 2008; where $\Gamma=1.43\pm0.18$ from $0.8-60$~\msun), NGC~602 (Schmazl et al. 2008, which is consistent with a Salpeter IMF from $1-40$\msun), and NGC~330 (Sirianni et al.~2002, consistent with Salpeter $>0.8$~\msun). 

Looking at young clusters and OB associations in the LMC/SMC, Massey~(2003) concluded that the massive star population in all observed regions is consistent with a Salpeter distribution.  He comments on the remarkably small scatter observed, considering the expected deviations due to the way that the data is processed (see also Ma\'{i}z Apell\'{a}niz~2008).   In the more remote regions of the LMC, and assuming a constant star formation rate over the past 10~Myr, Massey (2002) derived a much steeper $\Gamma=4\pm0.5$ IMF slope.  This led him to conclude that the IMF was environment dependent, in particular varying with density.  Similarly steep mass functions have been derived by Gouliermis, Brandner \& Henning~(2006) using HST WFPC2 data of a field near the bar of the LMC.   Parker et al.~(1998) used the Ultraviolet Imaging Telescope to derive the IMF of the LMC field and also found a steeper distribution (although not as extreme as the above studies) of $\Gamma = 1.8 \pm 0.09$ for the mass range of $7-35~M_{\odot}$.  Since the field is composed of stars of multiple ages, it is subject to the SFH-IMF degeneracy. Elmegreen \& Scalo~(2006) have shown that a decreasing star formation rate will be interpreted as a steeper $\Gamma$ if a constant star formation rate was assumed.  

Based on the above studies of clusters and associations the the Galaxy and Magellanic clouds, it already appears that strong variations in the high mass end of the IMF ($>1~M_{\odot}$) due to density ($35 < \rho_0$ [M$_{\odot}$ pc$^{-3}$]$ < 3 \times 10^4$)\footnote{$\rho_0$ is the central density of the cluster from Mackey \& Gilmore~(2003) for NGC~1818 and R136.  We note that the range in density would be even greater if OB associations were included.} and metallicity ($1/5~Z_{\odot} - 1~Z_{\odot}$) can effectively be ruled out.

\subsubsection{The Local Group}
\label{sec:localgroup}

Moving further out into the Local Group, we are naturally restricted to only probing the high-mass end of the IMF through resolved star counts.  M33, being a relatively face-on spiral, has offered the best laboratory to study the IMF beyond the Galaxy and the Magellanic Clouds.  Most studies to date have been based on giant H{\sc ii} regions associated with large OB associations.   Hunter et al.~(1996) found an index of the IMF of $\Gamma = 1.6 \pm 0.7$ for stars with masses between 6.5 and 18 \msun  ~based on HST imaging of NGC~604.  Similarly, Malumuth, Waller \& Parker~(1996) found $-1.3 \le \Gamma \le 1.0$ for stars more massive than 4~\msun\ in NGC~595.  These studies, like those carried out on populations in the LMC/SMC are subject to resolution/blending effects (e.g.~Ma\'{i}z Apell\'{a}niz~2008).

In addition to resolved star counts with high resolution imaging, numerous studies have used lower spatial resolution UV integrated spectroscopy of H{\sc ii} regions in M33 to constrain the initial mass function.   A discussion of this method will be given in \S~\ref{sec:uv-local}; here we just present results derived in the Local Group.  
Gonz{\'a}lez Delgado \& P{\'e}rez~(2000) found that the high mass end of the IMF in NGC~604 is consistent with a Salpeter IMF. Jamet et al.~(2004) found the same for NGC~588 based on integrated spectroscopy and photometric star counts.  Jamet et al. do warn, however, that stochastic sampling of the IMF can cause deviations in the integrated properties of the OB-associations, given their relatively low masses. This was also found by  Pellerin (2006) for five H{\sc ii} regions in M33, which were best fit with slightly flatter IMFs when fit with standard population synthesis models, but were consistent with a Salpeter slope when stochasticity was taken into account.  We will return to the use of UV spectral features to constrain the IMF in \S~\ref{sec:uv-local} and \S~\ref{sec:uv-highz} where these techniques have been applied to local and high redshift starburst galaxies, respectively. 

\subsection{Unresolved Stellar Populations}

\subsubsection{Super Star Clusters}
\label{sec:ssc}

Early results from the Hubble Space Telescope revealed partially resolved star clusters in starburst environments with masses and densities rivaling those of globular clusters (e.g. Holtzman et al.~1992).  These massive clusters have been referred to by many names: super star clusters (SSCs), young massive clusters (YMCs), starburst clusters and populous clusters, just to name a few.  Subsequent work has found such clusters in all star-forming environments, from relatively quiescent dwarf galaxies to Milky Way types (e.g. L$_{*}$ spirals)  and galactic mergers (e.g. Larsen~2006).  Due to their high luminosity, SSCs can be sampled  at great distances (and thus in a variety of galactic environments), providing valuable insights into potential environmental dependences.  We point the interested reader to Portegies Zwart, McMillan \& Gieles~(2010) for a review of the properties of these clusters.

The IMFs of these massive extragalactic clusters are particularly interesting as they represent the most extreme sites of star-formation, in terms of star-formation rate density, in the Universe.  For example, a massive cluster in NGC~1316 has an inferred star-formation rate surface density of  $\sim 5 \times 10^4~\msun~kpc^{-2}$~yr$^{-1}$ within its half light radius (Bastian et al.~2006; assuming it formed over a 3~Myr period), orders of magnitude larger than even the most extreme galaxy wide starbursts in the near or far Universe.  Even fairly common objects like R136 in the LMC attain  star formation rate densities of $\sim10^4~\msun~kpc^{-2}$~yr$^{-1}$, showing these "extreme objects" may in fact represent a wide-spread mode of star formation. 

Ho \& Fileppenko~(1996) suggested that dynamical light-to-mass ratios of young super-star clusters (SSCs)  could place important constraints on their IMFs.   Combining velocity dispersions ($\sigma_{\rm v}$), measured from high-resolution ground-based echelle spectra, with cluster half-light radii ($r_{\rm h}$), measured from HST imaging, enables the determination of cluster dynamical masses (\mdyn) through the application of the Virial Theorem.  This mass can then be compared to the mass derived by measuring the brightness of a cluster and applying age dependent L/M ratios from simple stellar population models (SSP), that adopt an IMF.  These stellar population--based masses will be referred to as \mpop.  Due to the difficulty of these measurements, this method is only sensitive to gross deviations from the nominal Kroupa/Chabrier-type IMF (e.g. if the clusters are highly deficient in low mass stars).

An early attempt to constrain the IMF of SSCs was performed by Sternberg~(1998), who used velocity dispersion measurements made by Ho \& Fileppenko (1996) to derive \lm\ for the young massive clusters NGC~1569A and NGC~1705-1.  Comparing the measured \lm\ to the fitted \lmpop ~ratios, Sternberg~(1998) concluded that the IMF of NGC~1569A was close to the Salpeter value, while NGC~1705-1 was depleted in low-mass stars (at the $1-2~\sigma$level).  

\begin{figure}[!h]	
\centerline{\psfig{figure=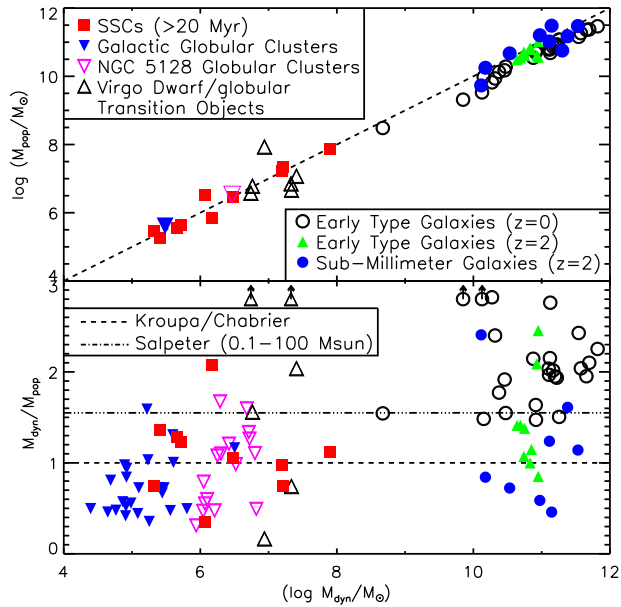,height=20pc}}
\caption{{\bf Top panel:} The measured dynamical mass {\mdyn} (or through Jeans dynamical modelling in the case of the Early Type Galaxies - see Cappellari et al.~2006) versus the stellar mass derived through modelling of their integrated light using simple/composite stellar population models which adopt an Kroupa type IMF.  The solid red squares are super star clusters with ages greater than 20~Myr (\S~\ref{sec:ssc}), the upside down blue filled triangle and open magenta upside down triangle represents the mean of 24 Galactic globular clusters and 16 globular clusters in NGC~5128, respectively (\S~\ref{sec:globulars}), open circles show Early Type Galaxies in the local Universe (\S~\ref{sec:local_dynamics}) and filled green triangles and blue circles show Early Type Galaxies and Sub-millimeter galaxies at high redshift (\S~\ref{sec:highz_dynamics}). {\bf Bottom panel:} The same as the top panel except now the ratio between \mdyn\ and \mpop\ is shown.  If the underlying IMF was well described by a Kroupa-type distribution a ratio of 1 is expected in this representation (shown as a dashed line).  If a Salpeter IMF (down to 0.1~\msun) is a good representation of the underlying IMF a value of 1.55 is expected (dash-dotted line).  Note that the galaxy points are upper limits, as a fraction of the \mdyn\ measurement is expected to come from dark matter.}
\label{fig:dynamics}
\end{figure}

A host of similar \lm\ studies have followed, with widely varying results: some analyses indicate SSC \lm\ values consistent with a standard Salpeter or Kroupa/Chabrier-type IMF (Larsen et al.~2004, Maraston et al.~2004), while others measure SSC \lm\ values that imply an over-abundance of low-mass stars (a so-called "bottom-heavy" IMF; Mengel et al 2002) or high-mass stars (a "top-heavy" IMF; Smith \& Gallagher~2001).  More worrisome is that some of these results have been derived within the same galaxy (e.g. M82, McCrady et al. 2003).  Bastian et al.~(2006) noted that there was a trend in how well cluster \lm\ ratios were fit by SSP models, in the sense that older clusters ($>20$~Myr) were well fit (with one exception to be discussed below) with a canonical IMF, while there exists a significant scatter in young clusters.  Additionally, they showed that NGC~1705 is only deviant by 1-2$\sigma$ from a Kroupa-type IMF.  The older clusters ($>20$~Myr) from the Bastian et al.~ sample are shown in Fig.~\ref{fig:dynamics} as solid red boxes (data taken from Bastian et al.~2006; Goodwin \& Bastian~2006).  In the upper panel we show the mass derived through dynamical measurements (\mdyn) vs. that derived using the observed luminosity and age-dependent mass-to-light ratios from SSP models (\mpop) that adopt a Kroupa  IMF.  In the bottom panel we show the \mdyn\ vs. the ratio \mdyn/\mpop\ where a value of 1 is expected if a Kroupa IMF is the underlying mass function and $\sim1.55$ if the IMF is better described by a Salpeter distribution.  The SSCs cluster around the expected value of a Kroupa distribution, and the variance is within the observational errors.

One cluster has received a particularly large amount of attention, the massive cluster known as 'F' in the starburst galaxy, M82.  Smith \& Gallagher~(2001) measured a \lm\ ratio for M82F, and found it was $\sim$ 3 times larger than the \lmpop\ ratio predicted for a population with a Kroupa IMF.   This measurement suggests M82F is highly deficient in low mass stars.  McCrady, Graham and Vacca~(2005) measured the velocity dispersion of the cluster in the near-IR and re-measured its size on HST High Resolution Camera imaging, and came to the same conclusion.  Based on a filter dependent size (i.e. the cluster appears larger in redder bands) these authors suggest that M82F may suffer from a high degree of mass segregation. If true, this would make the measured \mdyn\ a lower limit as the light is dominated by the most massive stars which would be preferentially found in the center, hence having a lower than expected velocity dispersion.  Bastian et al.~(2007) revisited M82F and found evidence for strong differential extinction across the face of the cluster ($\Delta A_{V} > 1$), and that a large H{\sc ii} region lies between the cluster and the observer, and suggest that both effects may complicate the measurements.  Even with these caveats, M82F represents the strongest case for IMF variations in SSCs.

The youngest clusters ($<20$~Myr) that have been analyzed show a much larger spread in their inferred \lm\ ratios, in particular having lower \lm\ ratios (by factors of $2-5$) compared to that expected from SSP models. If interpreted in terms of IMF variations, the result would be a "bottom heavy" IMF.  This is opposite to what is often reported in the literature, that starbursts are depleted in low-mass stars (see \S\S~\ref{sec:local_dynamics} \&~\ref{sec:cosmo}).  Additionally, these clusters would be more likely to survive the effects of mass loss due to stellar evolution, and hence we should observe older clusters with low \lm\ as well.  There are two alternative explanations for the high dynamical masses measured.  The first possibility is that removal of the cluster's natal gas has left it in a super-virial state for the subsequent $10-20$~Myr.  The cluster's velocity dispersion therefore has an additional expansion component, which will cause virial analyses to overestimate its mass  (Bastian \& Goodwin~2006; Goodwin \& Bastian~2006). 

The second possible explanation for the high \mdyn\ measurements of young clusters is that binary stars artificially broaden the velocity dispersion of a cluster.  If the binary fraction depends on stellar type, with more massive stars being exclusively in binaries (e.g.~Preibisch et al.~1999; Zinnecker \& Yorke~2007), this could cause young clusters to have low \lm\ ratios and older clusters to fit the SSP model predictions.  Preliminary estimates of the effect of binaries reported that this is unlikely to be affecting the clusters so far observed, as their velocity dispersion is dominated by the gravitational potential, due to their high masses (Kouwenhoven \& de Grijs~2008).  However, recently Gieles et al.~(2010) found that realistic mass dependent binary fractions could indeed affect the measured velocity dispersions of many of the young clusters in the sample discussed above. 

The integrated optical light from most young stellar populations is dominated by high-mass stars, making the detection of low mass stars difficult or impossible.  In extremely young clusters, however, low mass pre-main sequence stars are bright enough in the near-IR that their spectral features can be detected in high S/N spectra (Meyer \& Greissl~2005).  Quantitative analysis of these spectra can then place constraints on the ratio of high-to-low mass stars in the cluster.  This technique should be applicable for clusters young enough not to have evolved red giant stars (i.e. younger than 6-8~Myr) and is currently within the limits of 8-10m class telescopes for starburst clusters in the local Universe (Greissl et al.~2010), and further afield with the next generation of extremely large telescopes.

\subsubsection{The Upper Mass Cutoff and the IGIMF Theory}
\label{sec:igimf}

How large can the most massive stars be?  Does this vary with environment and what causes this limit?  From our perspective, an upper mass cut--off to the IMF is simply another parameter of the IMF that we might expect to vary as a function of initial conditions.  Zinnecker \& Yorke~(2007) have reviewed the literature of massive clusters ($> {\rm few} \times 10^{3}~\msun$) in the Galaxy/LMC and concluded that within these clusters an upper mass limit of $\sim150~\msun$ exists for the most massive stars.   The physics that imposes this cutoff is a critical, but separate issue; demonstrating that there is an upper mass limit can reveal some of the fundamental physics of star formation.  A question that we will now turn to is whether all clusters are subject to this same limit, or if lower mass clusters have a different upper limit to their mass function.

Summing the IMFs of a galaxy's constituent stellar populations (i.e. clusters, associations, and distributed field populations) produces an Integrated Galaxy Initial Mass Function (IGIMF; Weidner \& Kroupa~2005).  The IGIMF characterizes the bulk output of a galaxy's star formation activity, and provides an opportunity to investigate the sensitivity of the IMF on large-scale environmental effects.  Assuming that the stellar IMF is universal and sampled completely stochastically (i.e. a star forming in isolation is as likely to be an O star as a star forming in a much more populous region), then the IMF and IGIMF will be the same.  This is equivalent to saying that 100 clusters, each with a mass of 1000~\msun, will have a composite IMF which is indistinguishable from that of a single $10^5$~\msun\ cluster.  

The IMF will not be sampled completely in all cases, however, if the maximum mass of a star inside a cluster is proportional to the mass of the host cluster (e.g. Vanbeveren~1982; Weidner \& Kroupa~2006).  Stochastic sampling in clusters larger than few$\times10^4~\msun$ will fully populate the IMF up to an upper mass limit of 150 $\msun$, but lower mass clusters do not possess enough mass to fully sample the high mass end of the IMF (Bruzual \& Charlot 2003).   This should be the case if stars derive their masses from the resources in their surroundings (which they do): certain environments may simply not have enough material to form a high mass star.  Hence, there would be a lack of high mass stars in low star formation rate (SFR) galaxies, as these galaxies only form low mass clusters/associations (e.g.~Larsen~2006).  The high mass end of the IGIMF in high SFR galaxies would therefore be steeper than in low SFR galaxies, because there would be no high mass clusters capable of populating the highest mass end of the stellar IMF.   If true, this would have important effects on a wide range of astrophysical problems, most notably, on the inferred SFRs of (low SFR)  galaxies based on H$\alpha$~luminosities (e.g. Pflamm-Altenburg, Weidner \& Kroupa~2007; 2009)\footnote{This same effect would in principle be revealed by comparing 500 regions of 200 young stars to a 10$^5~\msun$ cluster, but in practice is almost impossible to observe.}.  

An analogy for this somewhat subtle, but important difference is the distribution of heights and wealth in villages and large cities (Clarke~2009).  On average, the tallest person in a village will be shorter than that in a large city, simply because the sample size is smaller.  There are no physical conditions restricting growth, so sampling statistics are the cause.  On the other hand, the wealthiest person in a large city is expected to be much richer than his/her equivalent in a village, and in this case sampling statistics  and an underlying physical cause (as the richest individual in a large city can draw from a larger economic base) are to blame.  Which of these comparisons is most applicable to how the IMF is sampled will have a strong impact on star/cluster formation theories.

Weidner \& Kroupa~(2006) have suggested that ``sorted-sampling" must be at play within young clusters (i.e. clusters form stars from low to high mass progressively).  Depending on the available gas reservoir (presumably related to the total emergent cluster mass) the mass of the most massive star will be limited as the IMF is sampled from bottom to top.  An argument in favor of such a scenario is that if high mass stars form first, the feedback caused by their ionizing radiation may effectively remove any remaining gas, hence stopping the formation of low mass stars.  Others argue that clusters are built entirely stochastically, as the relatively large (compared to the mass of individual stars) amount of molecular gas present in star-forming reservoirs allows high mass stars to form even in low SFR regimes (Elmegreen~2006)\footnote{The observations of young clusters are consistent with small age spreads though relative ages of high to low mass stars are notoriously difficult to assess (Hillenbrand 2009).}.  These two scenarios predict different relations between a cluster's total mass and the mass of its most massive star.  The recent compilation of Maschberger \& Clarke~(2008; see also Parker \& Goodwin~2007) of cluster masses and most massive stars includes several examples of low mass clusters containing high mass stars.  The authors note that previous compilations were biased by not including studies of massive stars surrounded by low-N clusters.  Their results tentatively suggest that random sampling is the preferred algorithm for forming clusters, in which case the IMF and IGIMF would be identical.  However, Weidner, Kroupa \& Bonnell~(2010) come to the opposite conclusion (that the most massive star does depend on the mass of the host cluster) based on a search of the current literature.  Determination of the cluster mass and the mass of the most massive stars (due to the presence of unresolved binaries) are difficult and potential stumbling blocks for all studies of this sort.  Clearly, this issue remains to be resolved.  

An additional component of the IGIMF model is the cluster/association initial mass function, which is also approximated as a power law with index $-\beta$, similar to Equation~\ref{eqn:powerlaw} ($N dM \propto M^{-\beta}dM$, where M is the mass of a cluster and N is the number of clusters with mass between M and M+dM).  The steeper the cluster initial mass function the more pronounced the difference between the IMF and the IGIMF, as one has fewer high mass clusters to form high mass stars.  Pflamm-Altenburg et al.~(2007) and Weidner \& Kroupa~(2006) adopt $\beta = 2.2$, whereas direct measurements of the cluster mass function are generally shallower, having $\beta = 2.0$ (e.g.~de Grijs et al.~2003).  This rather small difference in $\beta$ has a significant effect on whether or not the IMF and IGIMF are the same (Kroupa \& Weidner~2003; Elmegreen~2006).

\begin{figure}[!ht]	
\centerline{\psfig{figure=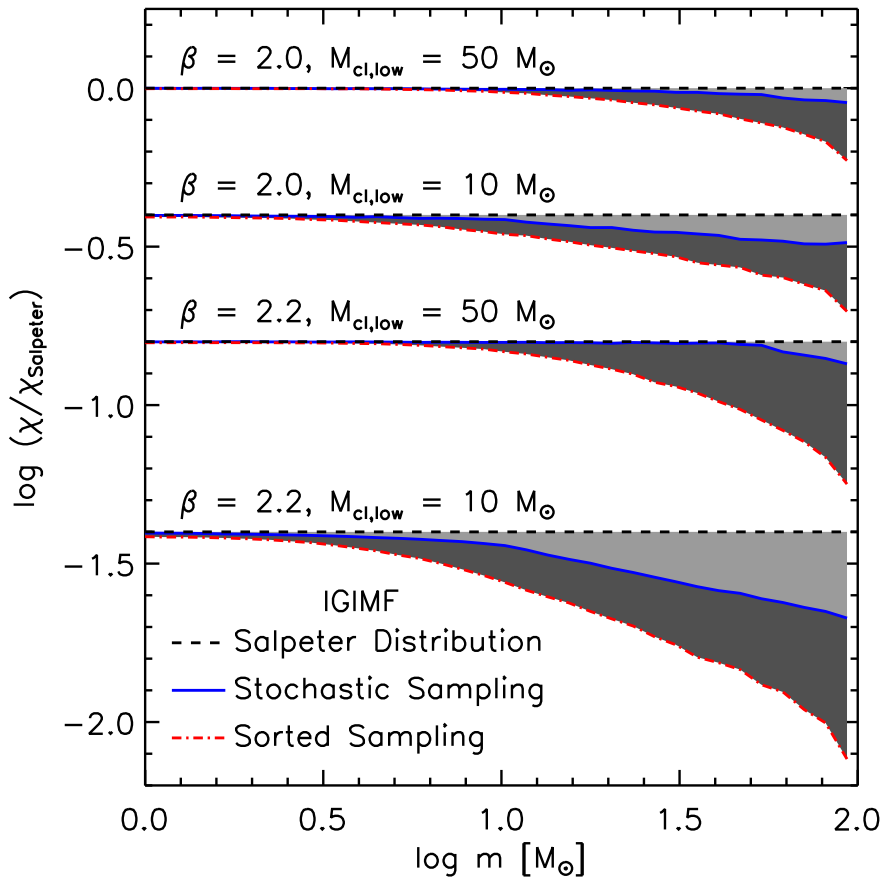,height=30pc}}
\caption{The difference between the IMF (drawn from a Salpeter distribution) and the Integrated Galactic IMF in the ``stochastic sampling" (Elmegreen~2006) and ``sorted sampling"  (Weidner \& Kroupa~2006) scenarios.  Shown is the fraction of stellar mass (see Equation~\ref{eqn:powerlaw}) in the IGIMF normalized to the expected mass (as a function of stellar mass) from the analytic Salpeter IMF (i.e. if the distribution is consistent with a Salpeter IMF the result is a flat line in this representation).  The various lines are for different assumptions on the underlying cluster initial mass function; $\beta$~ is the index of the power-law in the cluster mass function while $M_{\rm cl,low}$ is the adopted lower mass limit for clusters. The light shaded regions represent the relative lack of stars as a function of mass in the "stochastic sampling" scenario, while the dark shaded regions are the same, but for the ``sorted sampling" scenario. Since the fraction of high mass stars changes in each of the scenarios considered, the ratio of stellar mass dependent observations (e.g. H$\alpha$~flux or UV luminosity) will differ from the canonical values. (Adapted from Haas \& Anders~2010)}
\label{fig:igimf}
\end{figure}

In Fig.~\ref{fig:igimf} we show the difference between the input IMF and the resulting IGIMF for various assumptions on how stars are sampled from the underlying IMF and the mass function of the clusters/associations (different power-law indices and lower mass truncations).  Lada \& Lada~(2003) suggest a "turn-over" in the embedded cluster mass function at $\sim50~\msun$, although this value is somewhat uncertain due to selection effects.  As such, observations of clusters appear to prefer the upper most example in Fig.~\ref{fig:igimf}, which results in very similar IMF and IGIMF for any sampling scenario considered. The critical parameter is the ratio of $M_{\rm low}({\rm cluster})$ to $M_{\rm upper}({\rm star})$.

To test the validity of the IGIMF theory, several new, and
challenging, observations will be needed.  The first is the exact form of the relation between the mass of the most massive star and mass of the host cluster, for large samples of clusters as discussed by Maschberger \& Clarke~(2008).  The second is to understand in more detail the exact shape of the cluster mass function, in particular its index ($\beta$) and its lower mass limit (since $\beta \approx 2$ the cluster mass function diverges as the mass approaches 0, hence there must be a turnover at some lower mass).  
Finally, we need to determine the birth places for large
samples of high mass stars using kinematic studies to provide
a definitive answer whether they can form in isolation or
are always found associated with low mass stars.
We shall discuss other consequences of the IGIMF theory in the following sections.

\subsubsection{H$\alpha$, UV and Infrared Studies of Integrated Stellar Populations}

The UV, optical and infrared light of a galaxy is dominated by stars, either directly or through reprocessing of stellar light by dust and gas in the ISM.  In a typical stellar population comprised of stars with a range of ages, stars of different masses contribute differently to a galaxy's integrated light: far--UV emission is dominated by young ($<$10 Myrs) massive ($\geq15-20\msun$) stars which can ionize hydrogen, generating optical recombination lines such as $H\alpha$; near--UV emission is dominated by somewhat older ($<$100 Myrs), less massive ($\geq3$~\msun) stars; and optical broad--band colors are typically dominated (in quiescent galaxies) by even older solar--mass (0.6--3 M$_{\odot}$) stars (e.g.~Leitherer et al.~1999).   Comparing the relative strength of emission detected in these wavelength regimes can therefore provide an indirect constraint on galaxy-wide IMF variations.  An important caveat is that the IMF is not the only parameter that affects these observable properties; we must also account for the influence of metallicity, extinction, and the star-formation history of the galaxy.

Hoversten \& Glazebrook~(2008) estimated the index of the stellar mass function for $\sim10^5$ galaxies with SDSS photometry and spectroscopy, by comparing each galaxy's broadband SDSS colors with its H$\alpha$ equivalent width\footnote{A method first developed by  Kennicutt~(1983).}. They found a trend between the mass function index ($\Gamma$) of their best fitting model and the luminosity of the galaxy, in the sense that faint galaxies have steeper IMFs, galaxies similar to the Milky Way exhibit Salpeter slopes, and the most massive galaxies also have steeper than the Salpeter value.  This is qualitatively consistent with the predictions of the IGIMF theory. However, the stellar upper mass limit was not allowed to vary in their models.  These results depend on the assumed SFH, which in general is not precisely constrained with optical spectroscopy (e.g. Wild et al.~2009).  The ensemble is consistent with a $\Gamma = 1.5 \pm 0.1$ which is close to the Salpeter value. 

Several recent studies have used UV and H$\alpha$ measurements to characterize the high-mass IMF in external galaxies.  Meurer et al.~(2009) investigated a sample of galaxies with UV observations from the {\it GALEX} satellite and found a relation between the \hauv\ ratio and the H$\alpha$ surface brightness ($\Sigma_{H\alpha}$) of a galaxy.  This relation indicates that galaxies with low star-formation rates also possess low \hauv\-ratios, which the authors interpreted as a signature of a bottom-heavy IMF.  Subsequent studies by Lee et al.~(2009) and Hunter, Elmegreen \& Ludka~(2010) have confirmed this basic observational relationship between UV-to-H$\alpha$ luminosity ratio and star formation rate.

While these studies agree concerning the observed correlation between a galaxy's SFR and its \hauv\ ratio, they do not agree on the subsequent interpretation. Meurer et al. and Lee et al. show that part of the relation, though not all of it, can be explained by stochastic sampling, where low SFR galaxies are less likely to produce the highest mass stars, even if they were sampling from a universal IMF.   They also note that their results are well fit by the Weidner \& Kroupa~(2005) IGIMF theory (Pflamm-Altenburg, Weidner \& Kroupa~2009; see \S~\ref{sec:igimf}).   Meurer et al. also found that the observed \hauv\ ratios could be explained by a population of post-starburst dwarf galaxies in their sample (although the authors note that this is unlikely as many galaxies would need to be observed in a post-starburst phase).  Lee et al. also investigated additional explanations for this observed correlation, such as internal dust attenuation, uncertainties in the stellar models used, metallicity effects, and ionizing photon loss (see also Elmegreen \& Hunter~2006).  Lee et al. conclude that stochasticity, SFH, and ionizing photon losses all contribute to the offset in the correct sense, but none are likely to produce a strong enough trend alone to match the observations.  Hunter, Elmegreen \& Ludka, by contrast, favor a scenario where many ionizing photons from OB stars leak out of small galaxies\footnote{In many small galaxies, the Str\"{o}mgren radius around a single O-star can be larger than the gaseous galactic scale height, hence leakage of ionizing photons is expected (Melena et al.~2009).}, which causes an underestimation of the emission measure. Finally, Boselli et al (2009) similarly demonstrated that derived \hauv\ ratios are very sensitive to the adopted extinction corrections and micro-star formation history (from a few to tens of Myr).  Analyzing a restricted sample of ``normal" (i.e. low inclination, non-AGN, non-starburst) galaxies, Boselli et al. find \hauv\ ratios consistent with Kroupa/Chabrier type high-mass IMFs, and limit potential high-mass (M $>  2 M_{\odot}$) IMF variations to $\Gamma = 1.35$ for massive galaxies, and $\Gamma= 1.6$ for low-luminosity systems.  

 It remains to be seen if the H$\alpha$ deficiency noted above is truly indicative of a correlation between the shape of the IMF and SFR.  Future resolved color-magnitude studies of these galaxies, along with deep H$\alpha$ imaging of their halos will be necessary to settle this issue.

In a similar vein, Gogarten et al.~(2009) have studied individual UV-bright regions in the outer disk of the spiral galaxy M81 using high resolution HST imaging which can resolve massive stars.  They found that regions that lack H$\alpha$ do lack high mass stars, but that this is consistent with expectations of evolutionary fading, i.e. that as a stellar population ages the high mass stars that drive H$\alpha$ emission die first, leaving the region UV-bright but H$\alpha$ faint.  Hence, this particular ratio is highly sensitive to the SFH of a region.  If each region is formed in an instantaneous burst, and assuming that the number of regions formed per unit time is roughly constant, population models can be constructed to predict the number of UV-bright regions that will be H$\alpha$ bright or faint.  Zaritsky \& Christlien~(2007) have made such a model and found that this ratio will approximate the ratio between the lifetimes of stars that can produce H$\alpha$ to those that produce UV, i.e. 16~Myr/100~Myr.  Large studies of UV-bright regions in the outer disks of galaxies are necessary to confirm the predicted statistics.

Following on the Gogarten et al. result, Goddard, Kennicutt \& Ryan-Weber~(2010) have studied star forming regions in a sample of 21 galaxies, focussing on H{\sc ii} regions well beyond the optical radius of the galaxy.  Based on the H$\alpha$/UV ratio and statistical properties of the H{\sc ii} regions they do not find any evidence for variations in the IMF at the high mass end, once stochasticity is taken into account.  

These UV/H${\alpha}$ diagnostics appear sufficiently sensitive to a galaxy's recent SFH and/or radiative transfer effects that we do not yet consider them to be strong evidence for IMF variations.  Detailed studies of the selection effects inherent in the sample and the SFHs of the deviant galaxies using resolved stellar population studies, along with deep H$\alpha$ imaging of their halos, are needed to confirm that the IGIMF and IMF are truly different.  If confirmed, this would most likely indicate that the ratio of intermediate to high mass stars can vary with environment.

Finally, while the low mass end of the IMF does not significantly contribute the infrared spectral energy distribution of starburst galaxies, it does influence the absolute value of the SFR derived from e.g. emission lines (\S~\ref{sec:intro_applications}).  Emission line luminosities imply a specific number of high-mass stars, but converting this measurement into an estimate of the total stellar mass in a galaxy requires accounting for lower mass stars by adopting and integrating over a normalized IMF.  An IMF with many low-mass stars will imply a larger SFR for a given emission line luminosity than an IMF with fewer low-mass stars.  For a given reservoir of gas, these different SFRs also imply distinct gas consumption timescales: IMFs with more low-mass stars (and thus higher SFRs) imply shorter gas depletion timescales.   Goldader et al.~(1997) have compared the derived SFRs with the gas consumption timescale in a sample of 12 Ultra-luminous Infrared Galaxies.  They found that a Salpeter IMF (or steeper) that continues down to $0.1~\msun$ results in gas consumption timescales that are shorter than the mean age of the stellar population (derived from near infrared spectroscopy).  IMFs that ``turn-over" below $1~\msun$ (such as Kroupa/Chabrier types) and have a Salpeter slope above this value are consistent with their observations.

\subsubsection{UV spectral diagnostics}
\label{sec:uv-local}

The UV spectra of high mass stars ($>15$~\msun) possess broad emission and absorption lines,  formed in the winds these stars drive.  One approach is to study the line profiles of  C~{\sc iv}~1550\AA\ and S~{\sc iv}~1400\AA, that are heavily influenced by the shape of the high-mass stellar IMF, in spectra of unresolved stellar populations.  This method is, however, completely insensitive to variations in the low-mass regime.  The benefit of this technique is that it allows for a direct comparison of the upper end of the IMF in different environments, as it can be applied to objects as disparate as local starburst galaxies, quiescent resolved and semi-resolved star-forming regions in the Galaxy and LMC/SMC, and even to high redshift galaxies.  The latter will be discussed in \S~\ref{sec:uv-highz}, and a series of studies applying this diagnostic to giant H{\sc ii} regions in M33 were presented earlier in \S~\ref{sec:localgroup}.  These studies found that the observations of M33 were consistent with a Salpeter IMF in the high-mass end, but that stochastic sampling was important to take into account.  This caveat should have a much smaller effect on studies of strongly star forming regions or full galaxies which are expected to have fully sampled IMFs.  An additional concern is that UV spectral diagnostics depend on the assumed SFH of the region.  Most studies of full galaxies assume a constant SFH for $>20$~Myr, while UV studies of stellar clusters assume an instantaneous burst.

In one of the first studies using UV line profile diagnostics, Conti, Leitherer \& Vacca (1996) obtained HST STIS spectroscopy of the starburst galaxy NGC~1741.  Comparing the profiles of C~{\sc iv} and S~{\sc iv} they concluded that the high mass IMF in this galaxy is consistent with a Salpeter slope.  These and additional UV spectral diagnostics have been extensively modelled for starburst galaxies\footnote{Starburst galaxies are dominated by young stellar populations so that the star formation history is less uncertain.} using population synthesis, and their dependence on the burst history and metallicity are now relatively well known (e.g. Gonzalez Delgado et al.~1997; Leitherer et al.~1999; Maraston et al.~2009).  Nearby starburst galaxies also have UV spectral diagnostics that are consistent with a Salpeter distribution of high mass stars (see reviews Pellerin \& Robert~2007 and Leitherer~2009).  

Tremonti et al.~(2001) have studied the UV spectral properties across the actively star forming galaxy NGC~5253, focussing mainly on the difference between the spectra of ``clusters" and ``the field".  While this distinction was somewhat arbitrary (based on surface brightness), the cluster spectra were consistent with a Salpeter IMF extending up to $\sim100$~\msun, while the field appeared to be deficient in stars with masses above $\sim30$~\msun.  One interpretation is that the IMF in these environments is different.  Instead, the authors suggest that the lack of high mass spectral features in the field is due to an age difference between the young clusters and the field (cf. Massey~2002).  Stars will diffuse out from their natal unbound associations into the field over some timescale that depends on their initial velocity dispersion (e.g. Goodwin \& Bastian~2006), but is thought to be on the order of several Myr.  The most massive stars will therefore end their lives as supernovae before they reach the field.  Chandar et al.~(2005) have repeated this analysis for 12 local starburst galaxies and come to similar conclusions.

\subsubsection{Dynamics of Galaxies}
\label{sec:local_dynamics}

The \lm\ diagnostic introduced earlier as a constraint on the IMF of young massive clusters (see \S 3.2.1) is also useful in the study of whole galaxies.  As discussed above, this diagnostic cannot probe the shape of the IMF in detail; rather, they provide a single \lm\ value that can be compared to predictions of stellar population models generated by adopting distinctly different underlying IMFs.  The complications of using this method for galaxies rather than star clusters are: 1) galaxies are not simple stellar populations, so one needs to work out their star formation histories and often account for differential extinction and 2) the presence of additional dynamical components, namely gas and dark matter.

M82 is a nearby prototypical starburst galaxy, and as such, has often been used for IMF studies.  Unfortunately, due to its dusty nature and edge on orientation, extinction effects can be particularly troublesome.  Rieke et al.~(1993) modelled M82 and and found that the galaxy's K-band luminosity, Type II supernovae rate, CO index and kinematic mass were best reproduced by models which were deficient in low mass stars relative to the local IMF (compared to the Miller \& Scalo~1979 and Scalo~1986).  Subsequent studies by Satyapal et al.~(1995), however, concluded that the inner regions of M82 are fainter than previously derived as they adopt lower values for the extinction.  Based on these observations, they concluded that a ``normal" (i.e. Salpeter or Kroupa/Chabrier type) IMF is consistent with the data for M82.  More recently, detailed modelling of near to mid-infrared spatially resolved spectra indicate that the IMF may indeed flatten  below a few solar masses (F{\"o}rster Schreiber et al.~2003), i.e. the characteristic mass of the IMF would be 1-3~\msun, hence still suggestive of IMF variations, but less extreme than previous studies.  As discussed in \S~\ref{sec:ssc}, dynamical measurements of massive star clusters in M82 indicate ``normal" IMFs for some, although M82F appears to be deficient in low mass stars.  

For a fixed dynamical mass, the ratio of high to low mass stars can be inferred from estimates of the ionizing flux from high mass stars. It is worth noting that the Miller-Scalo (1979) and Scalo (1986) IMFs are in fact ``bottom-heavy" with respect to a Salpeter or  Kroupa/Chabrier IMFs due the steepness of the former at the highest masses.  Thus readers are encouraged to examine quantitatively results from the literature to determine whether a region is ``top-heavy", ``bottom-heavy", or even ``bottom-light" with respect
to modern estimates of the field star IMF.

In a similar study to that of M82, Doyon, Joseph, \& Wright~(1994) constructed models of the interacting starburst galaxy NGC~3256 in order to constrain the IMF.  By adopting different forms of the IMF and different star formation histories (constant, bursts, and declining), they matched their models against five observables (the strength of the CO bandhead, the equivalent width of Br$\gamma$, the fraction of Lyman continuum photons to infrared luminosity, emission line ratios, and total K-band luminosity).  The best fit models were deficient in low mass stars, however the IMF was found to be degenerate with the star formation history and also the assumed mass of the burst. Salpeter distributions in the high mass regime that turn-over at lower masses (i.e. Kroupa/Chabrier-type IMFs) were broadly consistent with the data.

Bell \& de Jong~(2001) also used dynamical diagnostics to investigate possible IMF variations in spiral galaxies.  For a galaxy with a given color and rotation curve, a maximal disk \lm\ ratio can be calculated by assuming that baryons make up as much of the mass of the disk as possible, and thereby minimizing the disk's dark component.  Comparing these maximal disk M/L ratios to those predicted by spectrophotometric spiral galaxy evolution models, Bell \& de Jong found that a Salpeter IMF overestimates the number of low mass stars in their disk galaxy sample, whereas a Kroupa IMF fits the data well.  They further use the scatter observed in the Tully-Fisher relation to place strict constraints on IMF variations between spiral galaxies.  Assuming that all of the scatter is due to IMF variations (i.e. assuming zero measurement error or variations in the star-formation history of the galaxies) implies a scatter in the stellar mass associated with a given luminosity less than a factor of two (see also McGaugh~2005).  

Along similar lines, Cappellari et al.~(2006) used a sample of 25 E/S0 galaxies with 2D integral field spectroscopy from the SAURON project to estimate the mass-to-light ratio of each galaxy (where the mass has been determined dynamically) and compared this to the expected mass-to-light of the stellar populations, as derived through spectral fitting.  They conclude that a Kroupa IMF is consistent with the data, and that there is no evidence for a varying IMF within these galaxies.  Their sample (shown as open black circles in Fig.~\ref{fig:dynamics}) comprises galaxies with luminosity weighted ages of between $2$ and $17$~Gyr, suggesting that the IMF does not vary with cosmic age.  Note, however, that due to the large mass in stellar remnants in old ($>10$~Gyr) stellar populations it is difficult to unambiguously differentiate between bottom light and ``normal" IMFs in these systems (see van Dokkum~2008).  We will return to this point in \S~\ref{sec:cosmo}.

\subsubsection{Chemical Evolution of Galaxies}
\label{sec:chemical}

One of the long lasting effects of the shape of the IMF on the observed properties of galaxies lies in the abundances of elements in stars.   Since different elements are thought to be produced by stars of differing masses, chemical evolution models developed to understand the abundance ratios and gradients in galaxies can also place constraints on the IMF.  The basic model parameters are: stellar yields (outlining which elements are produced by stars of different masses, and in what amounts), gas inflow and outflow (the inflow of ``primordial" material into the galaxy, and the loss of metal enriched material into the IGM), the SFH of the galaxy, and finally, the IMF.  These models can then be compared to the abundance and color gradients observed in actual galaxies.  The limiting factor for the accuracy of these models is the prescription adopted to describe stellar yields (including supernovae and supernovae rates), for which models can vary widely (e.g.~Chiappini et al.~2003).

 $\alpha$-elements (such as O, Mg, Ca, \&Ti) are thought to be produced by core collapse supernovae (SNe) at the end of a massive star's life.  Iron, on the other hand, is also produced in large amounts in Type Ia SNe, hence there is a delay  between the star-forming event and when Fe is released.  Since the $\alpha$-elements are thought to be produced in different amounts in a supernova, depending on the initial mass of the star, any variance in their ratios suggests a change in the IMF (Nissen et al.~1994).  The halo of the Galaxy exhibits a plateau at low metallicities in terms of the abundance ratios of $\alpha$-elements compared to iron below $[Fe/H] \sim -1.8$ (e.g.~Garc{\'{\i}}a P{\'e}rez et al.~2006), above this metallicity the [$\alpha$/Fe] abundance falls due to the production of Fe in Type Ia SNe.  The existence of this plateau suggests that these stars were formed out of material that was well mixed and of a similar composition.  This in turn suggests that the IMF of the stars in the generation before the observed halo stars did not vary significantly.  
 
Since the $\alpha$-element ratios appear to be constant down to the lowest metallicities probed, this suggests that (the high mass end) of the IMF has been invariant since a redshift of $z\sim3-5$.   This is supported by $\alpha$-element abundances measured in high redshift Damped Lyman-$\alpha$ systems (Molaro et al. 2001; Pettini et al.~2008) and in absorption systems along quasar sight lines (Becker et al.~2006), both of which are consistent with yields expected from a Salpeter IMF at the high mass end.  One caveat to this analysis is that it assumes that the gas that formed stars was enriched with homogeneously mixed elements produced in SNe.  Small systems, such as dwarf galaxies may have only experienced a handful of SNe at early times, so their $\alpha$-element abundances may be different than that seen in the Galactic halo (Venn \& Hill~2008; although see also Frebel, Kirby, \& Simon~2010).  

One can combine the relatively straightforward models above with different star-formation histories and stellar yields in order to model full galaxies.  Models of elliptical galaxies have been performed which can reproduce their photochemical properties (metallicity, gradients, etc) using a Salpeter IMF, which continues down to 0.1\msun\ (e.g. Pipino \& Matteucci 2004; Calura, Pipino \& Matteucci~2008).  These same models indicate that in disk galaxies, and in the solar neighborhood, a Salpeter IMF overestimates the metal production (Romano et al.~2005), implying an environmental dependent form of the IMF.  Similar models carried out by Nagashima et al.~(2005), however, directly conflict with these results and suggest that only extreme top-heavy IMFs can explain the properties of elliptical galaxies.

Ballero, Kroupa \& Matteucci~(2007) applied chemical evolution models to explain the metallicity distribution of stars in the Galactic bulge and M31, and conclude that the bulge IMF must have been different than the Galactic disk IMF.   These evolutionary models are subject to uncertainties related to the specifics of galaxy formation (e.g. Pipino \& Matteucii~2008) and the detailed treatment of gas inflow and outflow, both of which are not precisely known at the present time.  These uncertainties make it difficult to draw strong conclusions, particularly in the face of conflicting results based on direct star counts (e.g. Zoccali et al.~2000, discussed in \S 4.2.2).

Avoiding the complications of inflow and outflow, Renzini~(2005) reports that the metal content of galaxy clusters (a combination of the IGM material and galaxies) is well reproduced by a Salpeter IMF between $1$ and $25$\msun.  Conversely, Portinari et al.~(2004) have argued that all standard IMFs above 1\msun\ (e.g. Salpeter or Kroupa/Chabrier) fail to reproduce the metal content of the IGM once metals ``locked up" in low mass stars are taken into account.

While chemical evolution models have obtained notable success in explaining many abundance ratios, gradients and age-metallicity relations, significant uncertainties remain, and they provide conflicting constraints on galactic IMFs.  To our knowledge, no study has turned the argument around; assuming an invariant IMF, what constraints can be placed on the stellar yields, SFH of galaxies, galaxy build-up, and the progenitors of Type Ia and II supernovae?

There is observational evidence for a high mass biased IMF of primordial stars, in the form of the metallicity distribution function of stars in the Galactic halo.  The  lack of many extremely metal poor stars, $[Fe/H] < -4$, indicates that extremely low metallicity stars had lifetimes shorter than the current age of the Universe, implying that no low mass ($<1\msun$)  primordial stars were formed (e.g. Tumlinson~2006).  For a review of the observations of very low metallicity stars we refer the reader to Beers \& Christlieb~(2005).

A lasting effect of the first generation of stars is their production and distribution into the early Universe of heavy elements.  Which elements are produced and at what ratios depends on the initial masses of the primordial stars; if the stars are more massive than $\sim300$\msun, however, the entire core collapses into a black hole and no heavy elements are ejected (Heger \& Woosely~2002).   Hence, it is possible to place constraints on the primordial IMF from abundance patterns observed in low metallicity stars.  Tumlinson, Venkatesan \& Shull~(2004) attempt such a study and conclude that the first generation of stars need not be extremely massive (hundreds of solar masses) but could instead lie within the $8-40$\msun\ regime, with few or no low-mass stars.  It does appear, however, that the stellar IMF was significantly more top heavy in the early Universe relative to the present day. The exact transition period from high-mass biased to the current form remains elusive, although models, such as those presented in Smith et al.~(2009), are beginning to make detailed predictions of when and where this transition takes place.

\section{Cosmological IMF Variations}
\label{sec:cosmo}

A flurry of work in recent years has attempted to constrain variations in the stellar IMF over cosmological timescales and distances.  While most diagnostics are necessarily indirect, some offer intriguing evidence for variations, while others favor the standard, or ``universal" IMF.  In this section we analyze the evidence for and against IMF variations from studies aimed at medium to high redshifts.  Additionally, we examine local stellar structures (e.g. globular clusters and dwarf spheroidal galaxies) that formed at high redshifts, as this ``near-field cosmology" gives us the most direct handle on IMF variations, at least in the low mass regime.

There are theoretical reasons to expect that the IMF may vary cosmologically.  If the ``turn-over" in the IMF, observed to be $\sim0.1-0.3$~\msun\ today, is related to the minimum temperature $T_{\rm min}$ of molecular clouds (as might be expected if star-formation is linked to dust temperature), then the point of the ``turn-over" may vary since  $T_{\rm min}$ is linked to the CMB temperature (Larson~2005).  The reason why such a link may exist is that collapsing gas is only able to fragment as long as the temperature can continue to decrease when the density increases.  The minimum achievable fragmentation mass is then given by the Jeans mass that corresponds to the density and temperature at which the temperature minimum occurs.  Below this mass the gas begins to reheat as the collapse continues.  Alternatively, Elmegreen, Klessen \& Wilson~(2008) have stressed that a temperature and density are needed to define a characteristic mass (i.e. the ``turn-over" mass) and that the density was likely higher in the past too.  If $T_{\rm min} \propto (1+z)$ and $\rho \propto (1+z)^3$ then the thermal Jeans mass is independent of redshift: $T_{\rm min}^{3/2} / \rho^{1/2}=$constant.

The possibility of a cosmologically varying IMF is intriguing, as locally (see above sections) the characteristic mass of the IMF does not appear to depend strongly on environmental conditions.  Dav\'{e}~(2008) notes that while no conclusive evidence for a cosmologically varying IMF exists, whenever detections are reported, they invariably point to the IMF being more top-heavy in the past than at present times.  However, some caution should be exerted, as summed up by the ``Dav\'{e} theorem" (R. Dav\'{e} priv. comm.): ``All problems in extragalactic astrophysics can be solved by an appropriate choice of the IMF".

\subsection{The Distant Universe}
\label{sec:distant-universe}

\subsubsection{Rest Frame UV spectroscopy}
\label{sec:uv-highz}

The most direct way to determine the shape of the IMF for high mass stars in high redshift galaxies is to use the rest-frame UV spectral features introduced in \S~\ref{sec:uv-local}; the use of this diagnostic in both redshift regimes allows a valuable direct comparison of the high mass IMF in the nearby and high redshift Universe.  As noted earlier, however, this diagnostic is somewhat sensitive to a galaxy's recent SFH, in addition to the slope of its IMF.

While high redshift galaxies are intrinsically faint, they are within reach of 8-10m class telescopes if they are gravitationally lensed.  Two such galaxies have been heavily observed, MS~1512-cB58 at a redshift of Z= 2.7276 and ``The Cosmic Horseshoe" at Z=2.38115.  Using the observed C~{\sc iv} P-Cygni profile, which is sensitive to the slope of the IMF above $\sim10$~\msun, and the presence of stars above 50~\msun, Pettini et al.~(2000) conclude that high mass slope of the IMF in MS~1512-cB58 is consistent with the Salpeter value.  Quider et al.~(2009) come to the same conclusion for ``The Cosmic Horseshoe".   Steidel et al.~(2004) used similar techniques on a somewhat closer galaxy, at Z=1.411, and also conclude that the slope of the IMF is consistent with the Salpeter value for high mass stars.

With the next generation of extremely large telescopes, detailed studies of the rest frame UV in a large variety (from starburst to quiescent) of galaxies out to high redshifts should become possible.  This may prove to be a rich line of investigation to see if and where the IMF may be different.

\subsubsection{Galaxy Scaling Relations and Dynamics}
\label{sec:highz_dynamics}

By comparing the rest-frame color evolution of galaxies in clusters to their luminosity evolution, van Dokkum~(2008) argued for an IMF that evolves with redshift, with the IMF being much more top-heavy at higher redshift.  This work is based on the Coma galaxy cluster and seven other galaxy clusters at redshifts between 0.176 and 0.837, where accurate photometry and dynamical masses are available.  By examining changes with redshift in the relationship between a galaxy's rest-frame $U-V$ colors and mass, it is clear that early type (i.e. elliptical) galaxies become bluer with increasing redshift.  This result agrees with what is predicted if early type galaxies formed at high redshift and have been evolving (more or less) passively ever since.  However, the exact rate at which they are evolving in color appears to be inconsistent with completely passive evolution if the underlying IMF follows a Salpeter distribution near $\sim 1$~\msun.  The observations can be brought into agreement with the passive evolution scenario if the IMF is allowed to vary near $\sim 1$~\msun.

Using the characteristic mass of a Chabrier IMF as a parameterization (\mhat), van Dokkum~(2008) found that \mhat\ varies by a factor of 20, between today and a redshift of 4.  This drastic IMF variation will leave an imprint on older stellar populations (such as globular clusters and dwarf spheroidal galaxies), and appears to be at odds with observations.  We will return to this point in \S~\ref{sec:near-field}.  The assumptions underlying this work are that the slope of the color-mass relation observed in Coma also applies to higher redshifts and does not vary between clusters, that the simple stellar population models are known to the required accuracy, and that the epoch of galaxy formation is known to sufficient accuracy.

For large samples of galaxies, there exists a relatively tight relation between a galaxy's stellar mass ($\mstar$) and its SFR.  As discussed in \S~\ref{sec:intro_applications}, the SFR inferred from H$\alpha$, UV or IR observations is heavily dependent on the IMF assumed; the \dave\ relation is therefore implicitly or explicitly dependent upon an assumed IMF.   The observed trend of \dave\  is well reproduced by simulations of galaxy formation (largely independent of modelling parameters), however the offset of this relation (i.e. the zero-point) in the observations and models differ, with models over-predicting the total amount of mass formed for a given SFR over time.  The magnitude of this difference grows with increasing redshift, which led Dav\'{e}~(2008) to speculate that the characteristic mass, \mhat,  of the IMF may vary with redshift.  In order to reconcile the models with observations, he proposed that \mhat$=0.5(1+z)^2$, similar to that independently proposed by van Dokkum~(2008).  This implies a characteristic mass of 2, 4.5, and 8 \msun\ at redshifts of 1, 2, and 3, respectively.  As will be discussed in \S~\ref{sec:near-field}, this is inconsistent with observations of globular clusters and nearby dwarf galaxies, which have formation epochs between $z=3-5$.  Hence, either globular clusters and dwarf galaxies are not representative of the typical star forming event at these redshifts or systematics are at play in comparing the observed to modelled \dave\ relation.

Such an evolution also seems at odds with dynamical measurements of individual galaxies at z$\sim2$. Similar to studies described in \S~\ref{sec:local_dynamics}, Cappellari et al.~(2009) compared the mass derived for 9 galaxies at a redshift of $\sim2$ using Jeans dynamical modelling, Virial estimates, and stellar population models.  As the mass estimate based on stellar populations is largely dependent on the IMF adopted (see also \S~\ref{sec:ymcs}), this method can be used to test for a cosmologically varying IMF.  The authors find good agreement between the three mass estimates and conclude that the IMF in these galaxies is ``bottom light" (in their terminology), meaning consistent with a Kroupa/Chabrier type distribution.   Galaxies from their sample are shown as filled green triangles in Fig.~\ref{fig:dynamics}.  These conclusions are consistent with those presented by Cappellari et al.~(2006), discussed in \S~\ref{sec:local_dynamics}, where early type galaxies in the local universe were found to be consistent with Kroupa/Chabrier type IMFs.  Since these galaxies formed at high redshift ($z>3$, e.g. Renzini~2006) this argues against a cosmologically varying IMF.  Additionally, since the star-formation rates within these forming elliptical galaxies is expected to have been extremely high (e.g.~Genzel et al.~2003), this also argues against an IMF that correlates with star-formation rate.

Direct measurements of the M/L of high redshift sub-millimeter galaxies, which include measurements of the total stellar (through modelling of the observed spectral energy distribution) and gaseous mass (through CO measurements), are becoming possible (Tacconi et al.~2008).  Tacconi et al. compare the mass derived in stars and gas to that derived dynamically, through velocity dispersion/rotation measurements.  After correcting for the AGN component in the observed spectral energy distributions, Tacconi et al. find that the sub-millimeter galaxies in their sample are all consistent with a Chabrier IMF, assuming a fairly conservative CO-to-H2 conversion factor.  These points are shown as filled blue circles in Fig.~\ref{fig:dynamics}.

Elliptical galaxies exhibit a tight relation between their central velocity dispersion, effective radius, and central surface brightness (Djorgovski \& Davis~1987; Faber et al.~1987), which is known as the Fundamental Plane.  While the first two components (effective radius and central velocity dispersion) are structural properties, the central surface brightness depends on the luminosity of the galaxy, which in turn is dependent on the stellar population properties (age, IMF, etc). If the IMF varies as a function of galaxy mass then the Fundamental Plane will be affected, in particular it will "twist" when different redshifts are probed (see Renzini~2006 for a review).  This is because the IMF controls the rate of luminosity evolution, so if the IMF depended on galaxy mass, the Fundamental Plane would evolve at different rates for different high/low mass galaxies (Renzini \& Ciotti~1993).  Renzini~(2006) has searched for such an effect by comparing the Fundamental Plane of the Coma cluster in the local Universe, to galaxy clusters at redshifts of $z = 0.58$ and $z=0.83$.  He finds no systematic "twist" indicating that high and low mass elliptical galaxies have approximately the same IMF.

\subsubsection{Integrated Global Properties}
\label{sec:igp}

Hopkins \& Beacom~(2006) and Wilkins, Trentham \& Hopkins~(2008a) have compared the stellar mass density observed at a given redshift to the integral of the star-formation history (SFH) of the Universe up to that time.  These authors find that the integral of the SFH predicts higher stellar mass densities than observed.  By changing the slope of the IMF in the high mass star range (i.e. making it top heavy), the total mass formed from extrapolating an instantaneous indicator (like H$\alpha$) can be adjusted in order to bring the two measurements into agreement.  However, it should be noted that even at $z=0$ they require a non-standard index for the IMF to explain their observations, and that a more or less constant shift separates all data points (regardless of redshift) from the integrated SFH models, suggesting that systematics may be at play.   If an IMF is adopted which allows concordance at low redshifts, a cosmologically varying IMF must be invoked to explain observations for z$>0.7$ (Wilkins et al.~2008b). 

The results above depend on the accuracy with which we know the mass and star-formation rate density of the Universe at various redshifts.  These parameters are measured from galaxy luminosity functions, which are truncated by detection limits.  This truncation makes it necessary to extrapolate the galaxy luminosity function to low-luminosities before integrating to obtain the total mass and star-formation rate densities; the shape assumed for the luminosity function below the detection limit strongly affects the results of the integration. Using a large {\it UV} survey of Lyman Break Galaxies in the redshift range $z=2-3$, Reddy \& Steidel~(2009) find a much steeper luminosity function than earlier measurements, meaning that low luminosity galaxies make up a larger fraction of the mass density of the Universe at high redshift than previously assumed. Taking into account the low luminosity end of the {\it UV} galaxy distribution, the authors find that there is no need to invoke a varying or non-standard IMF to bring the integral of the star formation history of the universe into agreement with the observed stellar mass density .

\subsubsection{Galaxy number counts}

Sub-millimeter galaxies, detected in large numbers by the SCUBA detector on the JCMT, are thought to be massive starbursting galaxies, somewhat akin to local Ultra-Luminous Galaxies (e.g. Chapman et al.~2005).  Due to their brightness, large samples of sub-mm galaxies are being assembled for comparison with semi-analytic $\Lambda$CDM simulations.  

In order to bring the predicted numbers of faint sub-millimeter galaxies from semi-analytic $\Lambda$CDM simulations into agreement with observations, Baugh et al.~(2005) needed to assume an extremely top heavy initial mass function ($\Gamma=0$).  In their models, such an IMF both increases the total UV luminosity for a given mass, and increases the metal yield which in turn produces more dust.  This additional dust can absorb more UV photons, emitting them at IR wavelengths, and subsequently increasing the number of sub-millimeter galaxies observable.  Alternative models, which do not require top heavy IMFs, have also been put forward to explain the sub-millimeter galaxy counts (Granato et al.~2004).  

These semi-analytic models have a rather large number of free or semi-constrained variables, hence it appears premature to claim strong IMF variations before alternatives are investigated.

Using constraints on the total extragalactic background light, cosmic star formation history, and the present K-band luminosity density, Fardal et al.~(2007) conclude that a Salpeter IMF above 1~\msun\ cannot be fit to all three constraints.  The authors favor an evolving IMF, in particular a ``paunchy IMF" which is one with an excess of intermediate mass stars ($1.5-4\msun$).  Whether the cosmic SFH is known to the necessary degree (see \S~\ref{sec:igp}), other contaminating sources (e.g. AGN) can be accurately accounted for, and short bright stellar evolutionary stages (e.g. TPAGB stars) are known precisely enough in order to place strict constraints on the IMF and its evolution is questionable at the present time.

\subsection{Near Field Cosmology}
\label{sec:near-field}

While the most massive stars dominate the light that we see in star-forming galaxies at high redshift, it is the low mass stars which will remain for many Gyr, bearing the imprint of their IMF.  Locally, there are abundant examples of the remnants of star formation from the early Universe ($z \ge 3$), namely globular clusters (GCs), the Galactic bulge/halo and dwarf spheroidal galaxies.

\subsubsection{Globular Clusters}
\label{sec:globulars}

Globular clusters are thought to be, by and large, made up of a single population of stars with a common age and metallicity (although the discovery of composite populations in many GCs suggest a possibly more complicated formation process - c.f. Bedin et al.~2004).  GCs are some of the oldest objects in the Universe (e.g. Brodie \& Strader~2006), although GC-like objects continue to form in the present day (see \S~\ref{sec:ymcs}).  Due to the proximity of Galactic GCs, many studies of their mass functions have been carried out using deep HST and ground based imaging.  Fitting their  mass functions leads to characteristic masses (adopting the Chabrier form) of $\mc =0.33$~(Paresce \& de Marchi~2000).  This is similar, although slightly higher than that found for young clusters in the Milky Way disk (\mc$=0.1-0.3$, see \S~\ref{sec:pop1}).  However, as pointed out by van Dokkum~(2008), this is clearly much lower than predicted by the results of van Dokkum~(2008) and Dav\'{e}~(2008), which expect \mc$>4$\msun\ for the cluster formation epoch ($z=3-5$).  Additionally, intermediate age ($4.5$~Gyr, $z_{\rm form} \sim 0.4$) clusters in the SMC also appear to have mass functions consistent with globular clusters and young clusters (Rochau et al.~2007).

The similarity between the IMF of young clusters and globular clusters is shown in Fig.~\ref{fig:guido}.  Fitting tapered power-laws to the young and old clusters results in a consistent picture where the stars in both young and old clusters formed from the same underlying IMF (de Marchi, Paresce \& Portegies Zwart~2010).  However, the characteristic mass in older clusters does appear to be systematically larger than in young clusters and in the field (see \S~\ref{sec:field}).

Due to internal dynamical effects, GCs are expected to have their mass functions evolve with time. This has been observed in a relation between the slope of the present day mass function between $0.3-0.8$~\msun\ and the concentration (i.e. compactness) of a cluster (de Marchi, Paresce \& Pulone~2007).  The evolution of the stellar mass function within GCs in a tidal field has been studied through detailed N-body modelling (Baumgardt et al.~2008; see also Kruijssen~2009) who found that all of the GCs with well determined PDMFs could be explained with a Kroupa~(2002) IMF and subsequent dynamical evolution.  Their models show that \mc\ increases with age (specifically with time until disruption), hence the current value of \mc\ for GCs is in fact an upper limit for their initial values.  

Thus, if \mc\ was significantly different at a redshift of $3-5$ then the present, then globular clusters are not representative of the average star-formation at that epoch, as their stellar mass functions appear nearly identical those measured today in young star-forming regions in the Galaxy (see Fig.~\ref{fig:guido}).  An important caveat to the above analysis is that GCs with top heavy stellar IMFs (i.e. those with large values of \mc) would not have survived to the present time, as they would have long since dissolved (for a Kroupa or Salpeter type IMF the low mass stars provide most of the gravitational potential of the cluster).  Hence, it is possible that the current GCs in the Galaxy may not represent the average star-forming event at high redshifts. There are some chemical peculiarities, i.e. the fraction of C-rich stars, in the halo which indicate that this may have been the case (Tumlinson~2007).

Following on the discussion in \S~\ref{sec:ymcs}, we can also use the \lm\ ratio of globular clusters to test the stellar mass function within them.  In Fig.~\ref{fig:dynamics} we show the ratio for a sample of globular clusters of observed dynamical mass, M$_{\rm dyn}$ to the mass derived by measuring the luminosity and applying a mass-to-light ratio from IMF dependent simple stellar population models, $M_{\rm pop}$.   The sample shown consists of globular clusters in the Galaxy (McLaughlin \& van der Marel~2005; Kruijssen \& Mieske~2009) and NGC~5128 (Rejkuba et al.~2007; Kruijssen~2008).   Both samples are consistent with SSP model predictions for a Kroupa IMF, with the small offsets likely due to internal dynamical effects (Kruijssen~2009).  

\subsubsection{The Galactic Bulge and Halo}

The Galactic bulge is a largely coeval, old and metal rich stellar population that appears to be very similar in stellar content with other spiral bulges and elliptical galaxies (for a review see Renzini~1999).  Hence, the bulge represents a unique chance to study these populations through resolving individual stars.  Using HST NICMOS imaging of a region of the bulge, Zoccali et al.~(2000) demonstrate that the luminosity function of stars is in agreement with that observed for disk stars and globular clusters that appear not to have undergone large amounts of dynamical evolution, down to 0.15\msun.  


Gravitational microlensing studies also place constraints
on the bulge mass function, primarily from observations of
the distribution of event durations.  Alcock et al. (2000a)
describe the results of difference image analysis from the MACHO
data and conclude that the observations are consistent with
the present day mass function of Scalo (1986) with no evidence
for a large population of brown dwarfs.  Other microlensing
studies toward the LMC probe the mass function of compact
objects in the halo of the Milky Way.  Alcock et al. (2000b)
conclude that up to 20 \% in mass of a typical halo model
could be in the form of compact objects with likely masses between
0.15--0.9 M$_{\odot}$.  The suggestion that these might be very old
white dwarfs (e.g. Oppenheimer et al. 2001) would imply a mass function
for halo stars peaked at masses several times higher than found for the
galactic disk.  However, Reid (2005) argues strongly that there is no
convincing evidence for an anomalous population of white dwarfs that
could contribute significantly to the dark halo mass.
Like the globular clusters, no constraints can be placed on the
high mass end of the IMF, but below a few solar masses, the IMF
of old stellar systems and young star forming regions appears consistent.

\subsubsection{Nearby Dwarf Galaxies}

Dwarf spheroidal galaxies provide excellent laboratories in which to study the IMF while avoiding the dynamical evolution encountered by GCs. These galaxies have similar ages and star formation histories as GCs, however their stellar densities are orders of magnitude lower, so they experience correspondingly less dynamical evolution.  Wyse et al.~(2002) used HST data to construct a luminosity function (LF) of stars in Ursa Major, an old, local group dwarf spheroidal galaxy.  To avoid the complication of transforming observed luminosities to stellar mass, the authors directly compare the Ursa Major LF to that characterizing globular clusters of similar metallicity.  They find that the LFs are indistinguishable down to $\sim$0.3~\msun, and conclude that the IMF must be independent of density and the presence of dark matter into at least the sub-solar mass regime.

\section{Conclusions}

It was impossible to thoroughly review all observational results that bear on whether
the IMF is universal or not in the space (and time) available to us.  As mentioned in 
Section I, we have set out to review claims concerning variations in the IMF in order
to see whether other reasonable explanations of the results could be invoked without
resorting to a non--Universal IMF.  We do not find overwhelming evidence for large 
systematic variations in the IMF as a function of the initial conditions of star formation.  
We believe most reports of non-standard/varying IMFs have other plausible
explanations, but we have highlighted measurements that deserve significant 
follow-up to confirm their findings.  In this section we summarize the main points of the review and conclude with an outlook for future work.

\subsection{Synthesis of the Results Presented}

In the $>50$ years since Edwin Salpeter published his seminal work, it is remarkable that the ``best fit" value for the functional form of the stellar initial mass function above $1~\msun$ remains $\Gamma \simeq 1.3$.  Since that time, there has been tremendous work, especially on the sub-solar regime, and it appears that some consensus is arising within the community.  The general picture is that the high mass end is well approximated by a power-law with an index of $\sim-1.35$ (with an upper limit of $\simeq150~\msun$) that continues down to $\sim0.5-2\msun$.  At lower masses, the system IMF can be approximated as a log-normal
distribution, with a peak at $\sim0.2-0.3\msun$ and a dispersion of
$\sim0.5-0.6$\msun, or a series of broken power-laws with similar
shape.   The IMF in the sub-stellar regime is still rather uncertain
although many recent studies suggest that it is consistent with a
$\Gamma <\sim  -0.5$ power law and the extrapolation of the
log-normal distribution to lower masses.
 It remains to be seen whether a power--law departure of the IMF is required at the lowest masses and whether there exists a lower mass cutoff.  Below, we summarize important conclusions from each section of this review. 

\paragraph{Local clusters, associations, and the field} Locally, there does not appear to be any strong  systematic variation in the IMF.  Studies of many of the Galactic regions suffer from low number statistics, and as such, the tools to study them have been necessarily rather coarse.  However, when larger numbers of high mass stars do exist, a Salpeter form in the high mass end is often found.  At sub--solar masses, the remarkable similarity of the IMFs found in the field, dense massive clusters, and more diffuse low density star--forming regions, argues for a universal IMF above the hydrogen--burning limit.  Some nearby regions, in particular Taurus, exhibit stellar IMFs which appear to be inconsistent with those observed in other regions (e.g. Orion, IC 348, Cham I).  In a handful of other nearby regions (e.g. highlighted in the text), the most recent studies suggest deviations from the nominal IMF (Kroupa/Chabrier type), however detailed follow up observations are needed to confirm these reports.   

\paragraph{The low mass end} We are only just beginning to peer into the stellar/sub-stellar boundary of the IMF and large uncertainties remain.  
There is some evidence that the characteristic mass of a log--normal IMF in the Upper Sco region is significantly lower than found in the field or other regions.  If confirmed, Upper Sco would contain more brown dwarfs than other regions (like Orion).   Although there are some intriguing hints, current observations have not conclusively addressed whether there is a lower limit to the IMF, and if so, whether it might vary.

\paragraph{Extreme Galactic star-forming sites}  Large telescopes and advanced instrumentation allow us to probe the stellar content in some of the most extreme places of star-formation in the Galaxy.  Young massive clusters, like Westerlund~1, have power-law slopes that are consistent with the Salpeter value down to a few solar masses.  The Arches cluster, Quintuplet and NGC~3603 appear to have somewhat shallower indices, however the presence of mass segregation in these dense clusters complicates the analysis. The global IMF within these dense clusters appears to be largely consistent with a universal distribution.  The Galactic center has been extensively studied as the strong tidal field is (theoretically) expected to influence the IMF.  Studies have reported a range of IMFs in the region.   The detailed history of the stars found in the very center of the Galaxy (formed in-situ or having spiraled into their present locations) and the strong dynamical evolution both complicate the analysis.  Hints that the IMF may vary in these extreme environments require additional high resolution imaging and spectroscopy (which will be within reach with the next generation of telescopes) in order to be confirmed.

\paragraph{Nearby galaxies} The LMC and SMC provide a larger baseline to study the effects of environment on the IMF and their proximity allows the direct study of stars with masses above $\sim1~\msun$.  Studies of LMC/SMC clusters have been able to rule out strong variations of the IMF (above $1~\msun$) as a function of density (over 3 orders of magnitude in the cluster central density) and metallicity (between 1/5~\zsun\ and 1~\zsun).  Similar results have been presented for M33.  Integrated measurements (e.g. UV spectral diagnostics) of these regions will provide excellent templates in order to study starburst galaxies at mid-to-high redshift in order to directly compare the high mass end of the IMF.

\paragraph{Starburst galaxies} Studies of nearby (tens of Mpc) starburst galaxies and the star clusters within them have also led to a consensus that the IMF is similar to that observed in nearby star-forming regions.  This is particularly interesting for the stellar clusters, as their stellar densities (and star-formation rate densities) can vary over many orders of magnitude, implying that the IMF is not strongly influenced by the local environment.  The massive cluster, M82F, is still the best candidate to host a strongly deviant (top-heavy) IMF.

\paragraph{Integrated galactic IMF (IGIMF)}  The sum of the IMFs of all clusters, groups, and associations (clusters for short-hand) that are forming within a galaxy (the IGIMF) may not be statistically equivalent to the IMF of an individual region if the most massive star that can form in a cluster is dependent on the total mass of the cluster.  This has led to the development of different sampling algorithms for the build up of stars within clusters.  Some algorithms give a direct dependence of maximum stellar mass with group size, hence large populations of low mass clusters will never be able to form the highest mass stars.  If the mass distribution of clusters is known/assumed, models of the IGIMF and its properties can be developed.  If the cluster mass function is characterized by a power-law with index $-2$ (currently favored by observations), then the IGIMF and IMF are very similar for any sampling algorithm used.  For steeper cluster mass functions, the difference between the scenarios becomes more pronounced.  

\paragraph{The IGIMF and galactic properties}  Observations of low star-formation rate galaxies in the local universe show a diverging trend in their integrated H$\alpha$ to UV luminosity ratios.  Since H$\alpha$ and UV are due to stars of different masses, one interpretation of these results has been that the IMF is different in these galaxies, being deficient in high mass stars in small galaxies relative to massive galaxies.  However, an alternative hypothesis is that the H$\alpha$ flux is low within these galaxies because many of the ionizing photons from massive stars escape the galaxy without being absorbed by the interstellar medium.  Additionally, sample selection and ``bursty" star formation histories can also explain the observed trends without resorting to IMF variations.  Resolved stellar color-magnitude diagrams for these galaxies are required in order to definitively test which scenario is correct.  We note that the plateau in metallicity observed for the very lowest mass stars is consistent with theories describing the very first generation of star formation in the Universe. These theories predict that the first generation of stars should consist primarily of intermediate and high mass stars, with a radically different IMF than the standard IMFs discussed here (see Bromm \& Larson 2004).

\paragraph{Cosmological studies} Recent studies have also found evidence for a varying IMF with cosmological time, being weighted towards heavier characteristic masses at higher redshift.  Due to necessity, these analyses are indirect and hence are a ripe avenue for future research.  These reports have come from the evolution of the mass/color relation in galaxy clusters, the stellar mass - star-formation rate relation observed in large samples of galaxies, and in number counts of specific galaxies with respect to model predictions.  Each study has its own uncertainties and caveats, with some being dependent on model parameters.  Elemental abundances in high redshift star-forming galaxies, in Damped Lyman-$\alpha$ systems and along quasar lines of sight all point indirectly to a Salpeter-type IMF, at least for the massive stars.  Additionally, dynamical mass estimates of individual systems, along with stellar mass estimates from photometric modelling, appear to rule out strong IMF variations, with the handful of results published to date being consistent with a Kroupa/Chabrier-type IMF.  Reports that the stellar mass density did not equal the integral of the star-formation history of the universe, hence suggesting that star-formation rate indicators are not accurate at high-redshift, seem to have been resolved through pushing deeper into the galaxy luminosity function (i.e. the star-formation rate had been underestimated by counting galaxies and extrapolating to lower galaxy luminosities).

\paragraph{Near field cosmology}  We do have direct access to local environments that formed at high redshifts, namely globular clusters, the Galactic bulge/halo and local dwarf spheroidal galaxies.  In these stellar systems, the resolved stellar mass function appears to be consistent (at least in the low-mass end) with that observed in the local field population as well as in forming young clusters.  This argues against a cosmologically varying IMF.  

\subsection{Future avenues of study}

Future developments in instrumentation will enable more detailed probes of IMF variations than presently possible.  While large variations seem to be ruled out, more subtle systematic changes with initial conditions could lead to better constraints on theories of star-formation.  

For resolved stellar populations, large complete (or representative) samples of stars from well--defined
stellar populations in a range of environments are needed in order to test for potential IMF variations. Current observations of star clusters in local group galaxies are  confusion rather than sensitivity limited.  As such, the time to complete an observation (at a fixed signal--to--noise) goes as the diameter of the telescope to the power of six (in the diffraction-- and background--limited regime).  Ground--based adaptive optics on 6--10 meter telescopes as well as the new capabilities of WF3 on HST will continue to improve studies of the IMF in crowded fields for years to come, but future facilities will provide great advances.  The 6.5 meter James Webb Space Telescope, scheduled for launch in 2014, will conduct diffraction--limited observations in space from 1--25 microns, obtaining sensitivity to 1 M$_{\rm Jupiter}$ mass objects in the nearest youngest star--forming regions.  The next generation of adaptive optics assisted extremely large telescopes (ELTs) will be able to resolve individual stars within star-forming regions $< 1.0$ \msun\ out to 1 Mpc in the local group.  They will also allow us to push well into the sub-stellar regime in the inner and outer Milky Way as well as in the Magellanic Clouds.  

While crowding is not a concern for nearby stars, identification of nearby multiple systems will depend on continued advances in high spatial resolution, high contrast imaging.  Surveys of nearby stars sensitive to faint, close companions will enable a more complete characterization of multiplicity properties as a function of primary star mass, companion mass, and
orbital separation for comparison to the system IMF (Goodwin \& Kouwenhoven 2009). Detailed follow--up of newly identified multiple systems will also greatly improve empirical calibrations of the transformation between observables and theoretical estimates of mass (e.g. dynamical mass measurements). 

Aside from multiplicity surveys, studies of the Galactic field population will benefit most from advances in sensitivity and areal coverage.  A major breakthrough in understanding the field population of brown dwarfs over the lifetime of the Milky Way will come with the WISE satellite, launched in late 2009. This all--sky mid--IR survey should be able to detect the elusive ``Y dwarfs'' which represent the oldest, coldest sub--stellar objects in the Milky Way with temperatures below 400 K.  With an exquisitely sensitive sample of nearby, ultra-cool brown dwarfs, WISE will constrain the galactic field IMF well below the hydrogen--burning limit for a large fraction of the age of the galaxy (Wright 2008).  The SkyMapper survey will address the historical incompleteness in the nearby star sample at southern declinations, and synoptic surveys such as Pan--STARRS, the LSST, and the astrometric ESA GAIA mission will revolutionize kinematic studies of stellar populations throughout the galaxy.  We will be able to obtain parallaxes for stars and warm brown dwarfs to 100 pc or more, find complete samples of stars in young clusters and associations from their dynamics down to very low masses, and pinpoint the formation sites of massive (OB) stars across the Milky Way.  Do all high mass stars form in clusters? This observation will be critical to test the IGIMF scenario.

Continued development of long-wavelength observational capabilities promise to shed more light on the stages of star formation from which the IMF arises.  The Spitzer Space Telescope has helped to uncover hidden star formation in the densest molecular cloud cores (e.g. Evans et al. 2009), and future missions such as ESA's Herschel Space Telescope as well as millimeter interferometers, such as {\it SMA}, {\it CARMA}, {\it PdBI}, and {\it ALMA} allow us to peer into molecular cores of individual proto-stars refining our understanding of
how the molecular cloud core mass function is transformed into the initial mass function
of stars and sub--stellar objects (e.g. Myers 2009).  

Finally, the next generation of ELTs will also open up the high redshift universe to detailed studies.  Dynamical measurements of super star clusters out to 100s of Mpc and full galaxies spanning a large range in SFRs beyond $z\sim2$, will become possible providing a more direct handle on IMF variations.  Spatially resolved rest-frame UV spectroscopic diagnostics of star forming regions in high redshift galaxies will be in reach, which will allow a direct comparison with local templates; the recent revival of STIS and installation of COS will also enable more UV-based investigations of the high-mass IMF in nearby galaxies and clusters. 

Realizing the promise of these technical advances, however, will require a similar advance in the statistical analysis of IMF measurements.  As we enter this new era, we advocate a shift in the means used to characterize and search for variations in the IMFs of resolved stellar populations.  Specifically, we recommend that future IMF studies publish their derived space densities, such that IMF variations can be tested by using a direct statistical comparison of two measured IMFs, such as with a KS test, rather than by comparing the parameters of the analytic fit adopted to characterize these increasingly rich datasets.  If a functional form is fit to a IMF measurement, we suggest that statistical tools such as the F-test can provide quantitative guidance as to the most appropriate functional form to adopt, and that the uncertainties associated with the derived parameters be clearly reported.  By providing a more statistically sound basis for IMF comparisons, we will be better poised to uncover IMF variations where they do exist, and quantify the limits on IMF variations imposed by measurements consistent with a ``universal IMF".

\paragraph{Acknowledgments}

We would like to thank all of our colleagues for stimulating discussions and suggestions.  In particular we are grateful to Holger Baumgardt,  Jerome Bouvier, Ben Burningham, Giles Chabrier, Romeel Dav\'{e}, Bruce Elmegreen, Pavel Kroupa, Diederik Kruijssen, Kevin Luhman, Antonio Pipino, Niell Reid, Britton Smith, and Neil Trentham for detailed and insightful discussions that have significantly improved this manuscript.  We also thank Guido de Marchi and Marcel Haas for providing data ahead of publication.    Finally, we thank our better halves for their support, patience and understanding during the writing of this review, and A.B.M. for her help in enabling submission in this final form.

\end{document}